\documentclass[a4paper,11pt]{article}

\newcommand{\eg}{{e.\,g.}}
\newcommand{\ie}{{i.\,e.}}
\newcommand{\cf}{{cf.}}

\newcommand{\Tc}{T_\text{c}}

\usepackage{jheppub-priv} 

\usepackage{mathtools}
\usepackage{multirow}
\usepackage{amsmath,amssymb,amsfonts}

\usepackage{bm}
\usepackage{subcaption}
\usepackage[autostyle]{csquotes}

\usepackage{longtable}
\usepackage{booktabs}
\usepackage{hyperref}

\newcommand{\Phit}{\tilde{\Phi}}
\renewcommand{\Im}{\operatorname{Im}}
\renewcommand{\vec}[1]{\boldsymbol{#1}}
\renewcommand{\i}{\text{i}}
\newcommand{\del}{\partial}

\title{Dynamics of a Vortex Dipole in a Holographic Superfluid}

\author{Carlo Ewerz,}
\author{Andreas Samberg,}
\author{Paul Wittmer}

\affiliation{%
	Institut f\"ur Theoretische Physik,
	Ruprecht-Karls-Universit\"at Heidelberg,\\
	Philosophenweg~16,
	D-69120~Heidelberg, Germany\\
        and}
\affiliation{%
	ExtreMe Matter Institute EMMI,
	GSI Helmholtzzentrum f\"ur Schwerionenforschung,\\
	Planckstra{\ss}e~1,
	D-64291~Darmstadt, Germany}

\emailAdd{c.ewerz@thphys.uni-heidelberg.de}
\emailAdd{a.samberg@thphys.uni-heidelberg.de}
\emailAdd{p.wittmer@thphys.uni-heidelberg.de}

\abstract{
We use holography to investigate the dynamics of a vortex--anti-vortex 
dipole in a strongly coupled superfluid in 2+1 dimensions. The system
is evaluated in numerical real-time simulations in order to study the evolution
of the vortices as they approach and eventually annihilate each other. 
A tracking algorithm with sub-plaquette resolution is introduced which permits 
a high-precision determination of the vortex trajectories. With the increased precision of
the trajectories it becomes possible to directly compute the vortex velocities and accelerations. 
We find that in the holographic superfluid
the vortices follow universal trajectories independent of their initial separation,
indicating that a vortex--anti-vortex pair is fully characterized by its separation. 
Subtle non-universal effects in the vortex motion at early times of the evolution
can be fully attributed to artifacts due to the numerical initialization of the vortices. 
We also study the dependence of the dynamics on the temperature of the superfluid.
}

\dedicated{\center Dedicated to the memory of Steve Gubser}

\begin{document} 

\maketitle
\flushbottom

\section{Introduction}\label{Sec:Intro}

Experimental advances in studying the non-equilibrium dynamics of isolated quantum-many-body systems have in recent years led to the quest for a better theoretical understanding of such dynamics. 
Systems of interest in this context include ultracold Bose gases \cite{Anderson2001a,Eiermann2004a,Sadler2006a,Weller2008a.PRL101.130401,Weiler2008a,Neely2010a}, 
semi-conductor exciton–polariton 
superfluids \cite{Kasprzak2006a.etal,Lagoudakis2008a,Lagoudakis2009a,Amo2011a,roumpos2011single} and 
topological condensed matter systems \cite{Qi2009a}. 
The non-equilibrium dynamics of such systems can be observed after a quench into a far-from-equilibrium initial state, 
as it can be realized experimentally by a sudden change of a parameter of the system, \eg\ temperature or the scattering length of the constituent particles. 
During the relaxation process and before thermal equilibrium is reached the system's dynamics is generically strongly nonlinear and exhibits properties of quantum turbulence, the quantum analogue of classical turbulence.
This line of research adds new aspects to the general phenomena of quantum and classical turbulence which have attracted considerable attention since decades and include, among others, universal scaling of correlation functions \cite{Kolmogorov1941a,Kolmogorov1941b,Kolmogorov1941c,Obukhov1941} with characteristic energy cascades in the system, quasi-stationary flow \cite{Richardson1920}, and the observation of non-thermal fixed points \cite{Berges:2008wm,Berges:2008sr,Scheppach:2009wu,Berges:2010ez,Prufer:2018hto,Erne:2018gmz} (see \cite{Schmied:2018mte} for a review).

The strongly nonlinear nature of the dynamics poses a great challenge as it renders a theoretical description very intricate. 
Even if the system's constituents are weakly coupled, commonly used methods such as perturbation theory break down
as a consequence of strong correlations that build up dynamically, 
often related to the occurrence of highly nonlinear topological excitations in the system.
These include in particular quantized vortices, the quantum analogues of classical eddies.
In modern experiments with Bose--Einstein-condensed ultracold quantum gases 
vortex ensembles can even be produced by stirring the condensate with a
laser \cite{Raman2001a.PRL.87.210402R,Abo-Shaeer2001a}. 
Quantized vortices are known to play a crucial role also in superfluid helium \cite{Donnelly1991a} where strong
coupling prevails in the system and the theoretical description is even more challenging. 
In the present work we will be concerned with the dynamics of a vortex--anti-vortex pair
which is the most basic configuration that allows one to study the interaction of topological defects. 

We will use the theoretical framework of holography or gauge/gravity duality for our study.
It offers a promising pathway towards a better understanding
of non-equilibrium dynamics and is particularly well suited for the description of strongly coupled quantum systems. 
In holography, a strongly coupled quantum field theory without gravity in $d$ dimensions is mapped
to a theory with classical gravity in $d+1$ dimensions.
The quantum behavior of the field theory is thus encoded in a classical gravitational theory in a higher-dimensional spacetime. 
A variety of holographic dualities has been devised since such a duality was first proposed in the form
of the AdS/CFT correspondence \cite{Maldacena:1997re,Gubser:1998bc,Witten:1998qj}.
AdS/CFT posits that $\mathcal{N}=4$ super Yang--Mills (SYM) theory, a 
supersymmetric and conformal gauge theory (CFT) in $3+1$ dimensions, is dual to type IIB string theory 
on $\text{AdS}_5 \times S^5$, where AdS$_5$ stands for an Anti-de Sitter spacetime and $S^5$ denotes a five-sphere.
A key feature is that the AdS/CFT correspondence simplifies in the limit of strong coupling and large number of colors on the field theory side,
corresponding to which the string theory side simplifies to a weakly coupled theory of Einstein gravity.
Generalizations to dualities between other strongly coupled quantum theories and weakly coupled gravitational theories
have become a powerful tool to address many problems in strongly coupled and hence manifestly nonperturbative quantum field theories.
The quantum theory in $d$ dimensions is usually referred to as the `boundary theory' as one can imagine it to `live' on
the boundary of the $(d+1)$-dimensional AdS spacetime.  
The AdS spacetime away from the boundary, on the other hand, is referred to as the `bulk'.

Holographic dualities can in particular be used to describe quantum field theories at finite temperature. 
Here, typically, the dual gravitational representation includes a black hole in the bulk of the AdS space, with its
Hawking temperature being equal to the temperature in the boundary theory. Similarly, a chemical
potential in the field theory corresponds to a charged black hole in the bulk. 
Holography has been applied to a variety of physical systems (see \eg\ \cite{Adams:2012th} for a review), ranging 
from quark--gluon plasma physics, see \eg\ \cite{CasalderreySolana:2011us}, to condensed matter 
systems, see for instance \cite{Hartnoll:2016apf}, including the pioneering work on 
holographic superconductors and superfluids
in \cite{Gubser:2008px,Hartnoll:2008vx,Herzog:2008he} (see \cite{Herzog:2009xv} for a review). 

Specifically, it has been found \cite{Gubser:2008px,Hartnoll:2008vx,Herzog:2008he} that an
Abelian Higgs model coupled to gravity on a $(3+1)$-dimensional Anti-de Sitter spacetime with a black hole 
has a dual interpretation in terms of a superconductor or superfluid (for a review, see \cite{Hartnoll:2009sz}).
Superfluidity is associated with the breaking of a global U$(1)$ symmetry.
In the holographic model, this is incorporated by the breaking of a local U$(1)$ symmetry and the associated 
condensation of the scalar field in the bulk. 
Remarkably, the simple Abelian Higgs model with classical gravity encodes the quantum behavior 
of the superfluid, even if strong correlations are present in the system.
It is therefore well suited for studying the real-time dynamics of the superfluid far from equilibrium. 

Extensive analytical as well as numerical studies of the holographic superfluid have been performed. 
Among them are studies of the details of the phase transition into the superfluid 
phase\footnote{Some attention has also been paid to modified holographic superfluid models and how gravity modifications 
affect the nature of the phase transition, in particular with regard to critical exponents that deviate 
from the mean-field values they take for the original holographic superfluid,
\eg\ \cite{Franco:2009yz,Franco:2009if,Herzog:2010vz}.} \cite{Herzog:2008he,Maeda:2009wv}, computations of the 
speed of sound and the investigation of sound waves,
see \eg\ \cite{Herzog:2008he,Basu:2008bh,Herzog:2009ci,Yarom:2009uq,Amado:2009ts}, and studies of topological defects in the 
system, see \eg\ \cite{Keranen:2009vi,Keranen:2009ss,Keranen:2009re,Keranen:2010sx,Adams:2012pj,Dias:2013bwa,Chesler:2014gya,Ewerz:2014tua,Du:2014lwa,Lan:2016cgl,Lan:2017qxm,Guo:2018mip,Lan:2018llf,Li:2019swh,Xia:2019eje,Yang:2019ibe,Xu:2019msl}. 
With regard to topological defects, much attention has been devoted to the study of vortices and in particular  
ensembles of vortices and anti-vortices.
While quantized vortices constitute an intense field of research in general, they play a particular role in the holographic superfluid as
the main sites of energy and momentum dissipation in this system \cite{Adams:2012pj}. 
To wit, they have a geometric interpretation in that tubes emanating from the vortex cores at the boundary punch holes
through the scalar charge cloud in the bulk so that energy can flow from the boundary to the black hole where it is absorbed,
see \cite{Adams:2012pj}. 

In the present work, we will study the annihilation process of a single vortex--anti-vortex pair, or vortex dipole, in the holographic superfluid. 
By considering just one vortex and one anti-vortex we have a clean environment for investigating their mutual interaction. 
At the same time, vortex pairs can be viewed as basic building blocks for the interactions in larger ensembles of vortices
and are thus relevant in a more general context. Our method builds on previous work in \cite{Ewerz:2014tua} where the vortex dynamics of
an ensemble of vortices was studied for certain classes of quench-like initial conditions, and we use the same numerical
implementation.
While the annihilation process of a vortex dipole in a holographic superfluid has been studied before \cite{Lan:2018llf}, 
we are able to study new aspects of this system by using new methods. Specifically, by introducing 
an advanced tracking technique, we drastically increase the precision of the vortex trajectories
that we extract from the numerical simulations. This gain in precision allows us to compute directly
the vortex velocities and accelerations. We can then infer universal properties of the motion of the vortex dipole
and can cleanly identify a power law for the behavior of the vortex--anti-vortex separation.
We also compare the vortex trajectories to a general point-vortex model from the literature. 
Furthermore, we study the temperature dependence of the vortex dynamics.
Finally, we discuss some potential further applications of the methods for tracking vortices
developed here. 

Our paper is organized as follows. 
In section \ref{Sec:HoloFluid} we review the gravitational model for the holographic superfluid. We discuss
its equations of motion and their numerical implementation used for the real-time simulation of the system. 
Section \ref{Sec:VortAntiVort} is concerned with the study of the vortex--anti-vortex pair in the holographic superfluid.
We describe the numerical initialization of vortices and introduce a new method to determine their positions during
the evolution of the system with high precision. Subsequently, we present our results for the vortex trajectories
from the initial separation to the annihilation. Here we also discuss universal properties of the motion of the vortex dipole 
and compute velocities and accelerations of the vortices. 
In section \ref{Sec:MuDependence} we study the dependence of the vortex 
dynamics on the temperature of the superfluid. We close with a summary and outlook in section \ref{Sec:Summary}.
Some technical details are presented in the appendices. 

\section{Holographic Superfluid}
\label{Sec:HoloFluid}

In this section we review the $3+1$-dimensional gravitational model dual to a superfluid in $2+1$ dimensions.

\subsection{Holographic Model}
\label{Sec:Model}

In his seminal work \cite{Gubser:2008px}, Gubser showed that, if coupled to gravity with a negative cosmological constant,
even a simple Abelian Higgs model without self-interaction exhibits spontaneous symmetry breaking. 
This laid the ground for a holographic description of superconductivity and superfluidity as subsequently developed
in \cite{Hartnoll:2008vx,Herzog:2008he}.\footnote{The
  holographic superfluid was devised in \cite{Herzog:2008he} after the dual description of a superconductor
  was introduced and discussed in \cite{Hartnoll:2008vx}. From the field-theory perspective,
  the main difference between a superfluid and a superconductor is that for a superfluid a global symmetry is
  spontaneously broken whereas for a superconductor a local symmetry is spontaneously broken.
  The differences between the holographic superfluid and the holographic superconductor with regard
  to topological vortex defects are discussed in \cite{Dias:2013bwa}, for example.} 
The spontaneous breaking of the local $U(1)$ symmetry in the AdS bulk occurs due to the negative contribution of the gauge 
field to the effective scalar mass in the vicinity of the black-hole horizon which drives the system unstable and leads to the 
formation of a scalar charge cloud. In the dual field theory this corresponds to the breaking of a global $U(1)$ symmetry and
a phase transition from the normal fluid to a superfluid state. 
The gravity model holographically describing a $(2+1)$-dimensional superfluid is given by the action \cite{Gubser:2008px}
\begin{align}\label{action}
S=\frac{1}{2\kappa}\int\,\text{d}^4x\,\sqrt{-g}\left(  \mathcal{R}-2\Lambda+\frac{1}{q^2}\mathcal{L_\text{matter}}\right)\,,
\end{align}
where $g$ is the determinant of the metric $g_{MN}$ ($M,N=0,\dots,3$), $\mathcal{R}$ its associated Ricci scalar, 
$\kappa$ Newton's constant in $4$ spacetime dimensions,
and $\Lambda=-3/L^2_\text{AdS}$ the cosmological constant of the $(3+1)$-dimensional 
Anti-de Sitter spacetime with curvature radius $L_\text{AdS}$.  
The matter part $\mathcal{L_\text{matter}}$ of the Lagrangian is given by
\begin{align}
\mathcal{L_\text{matter}}=-\frac{1}{4}F_{MN}F^{MN}-|(\nabla_M- \text{i} A_M)\Phi|^2 -m^2|\Phi|^2\,,
\end{align}
where the charge $q$ of the scalar field has been pulled out of the Lagrangian $\mathcal{L_\text{matter}}$ by an appropriate rescaling of the fields. 
Using this convention makes evident that $q$ controls the effect of the matter part of the model on the dynamical gravity, \cf\ \eqref{action}.
The mass of the scalar field $\Phi$ is $m$.
We use $\nabla_M$ to denote the Levi-Civita connection of the metric. 
The gravity model is thus given by a simple Abelian Higgs model formulated on a $(3+1)$-dimensional 
gravity background with minimal coupling between the complex scalar field $\Phi$ and the gauge 
connection $A_M$. The kinetic term for the Abelian gauge field is formulated in terms of its field 
strength tensor $F_{MN}=\nabla_MA_N-\nabla_NA_M$. Note that no terms of higher order in the scalar field than the quadratic 
mass term are needed. This is different from the usual Higgs mechanism in the Standard Model 
where the $\Phi^4$ interaction term is essential for spontaneous symmetry breaking to occur. 
In contrast to flat Minkowski spacetime, the AdS-type spacetime allows for 
an instability to occur even if there is no self-interaction of the scalar field.
The squared scalar mass $m^2$ may even be negative without the symmetry being broken as 
long as it is above the Breitenlohner--Freedman bound \cite{Breitenlohner:1982bm,Breitenlohner:1982jf,Mezincescu:1984ev}.
At low temperatures, the effective scalar mass can be lowered below this bound by the negative
contribution of the bulk gauge field, thus giving rise to an instability and symmetry breaking. 
In this work we choose $m^2=-2/L^2_\text{AdS}$ which is above the Breitenlohner--Freedman bound,
and we set $L_\text{AdS}=1$ for convenience.

The interpretation of this model according to the holographic dictionary is the following. 
The Abelian gauge field in the bulk $A_M$ is dual to a global conserved $U(1)$ current operator $j^\mu$ in 
the boundary field theory ($\mu=0,1,2$). Similarly, the complex scalar field $\Phi$ with mass $m$ and charge $q$, to 
which the gauge field $A_M$ is dynamically coupled, is dual to a complex scalar field operator $\Psi$. 
This operator plays a crucial role as its quantum expectation value $\psi=\langle\Psi\rangle$ is the order 
parameter for the phase transition from a normal fluid to the superfluid. 
In a usual field-theory context, superfluidity is described by a complex scalar field whose vacuum expectation value  
assumes non-zero values in the symmetry-broken phase, corresponding to the existence of a superfluid condensate. 

Instead of considering the full set of equations of motion corresponding to the action \eqref{action}
we simplify the model by taking the so-called probe approximation. This means that the gravity part of the action
is solved independently of the matter fields, \ie\ the back-reaction of the matter fields on the metric is neglected. 
This approximation is justified if the scalar charge $q$ is large. 
The matter part of the model is then solved in the fixed gravity background. 
For a detailed discussion of the probe limit and its validity see \cite{Hartnoll:2008vx} and, in the context of
vortex defects, \cite{Albash:2009iq}.

The pure gravity part which comprises only the first two terms in the action \eqref{action}, \ie\ the 
Ricci scalar and the cosmological constant, can be solved straightforwardly. 
The vacuum Einstein equations are given by 
\begin{align}
\mathcal{R}_{MN}-\frac{1}{2}\mathcal{R}\,g_{MN}+\Lambda\,g_{MN}=0\,. 
\end{align}
They are solved by a $(3+1)$-dimensional Schwarzschild Anti-de Sitter spacetime that
has a planar black hole with horizon situated at the bulk position $z=z_\text{h}$.
In this work we choose to work with infalling Eddington--Finkelstein coordinates in which the line element is given by
\begin{align}\label{metric}
	\text{d}s^2=\frac{L_\text{AdS}^2}{z^2}\left(-h(z)\text{d}t^2+\text{d}\bm{x}^2-2\text{d}t\,\text{d}z\right)\,,
\end{align}
with the horizon function
\begin{align}
	h(z)=1-\left(\frac{z}{z_\text{h}}\right)^3\,.
\end{align}
The choice of infalling Eddington--Finkelstein coordinates is convenient for two reasons. 
Firstly, unlike the usual Schwarzschild coordinates they do not exhibit a coordinate 
singularity at the black-hole horizon. Secondly, all null geodesics falling into the black hole are of radial nature. 
These two features are particularly useful for numerical 
simulations in holography \cite {Chesler:2013lia} due to the ensuing natural implementation of boundary conditions.
In these coordinates, $z$ is the holographic (radial) coordinate and $(t, \bm{x})=(t, x, y)$ are 
the time and remaining spatial directions, which also make up the coordinates of the dual field 
theory that can be thought of as living on the boundary at $z=0$ of the Anti-de Sitter spacetime. 
As a radial coordinate $z$ is taken to be positive and one refers to the spacetime for $z>0$ as the bulk.
The Hawking temperature associated with the metric \eqref{metric} is 
\begin{align}\label{temp}
	T=\frac{3}{4\pi z_\text{h}}\,, 
\end{align}
and it coincides with the temperature of the boundary theory. 
Taking the background metric to be fixed one can vary the action \eqref{action} with 
respect to the matter fields to obtain their equations of motion,
\begin{align}
\nabla_MF^{MN}&=J^N\,,\label{Maxwell}\\ 
\left(-D^2+m^2\right)\Phi&=0\,, \label{KGE}
\end{align}
with the current 
\begin{align}\label{current}
	J^N&=\text{i}\left(\Phi^\ast D^N\Phi-\Phi\left(D^N\Phi\right)^\ast\right)\,,
\end{align}
and the general covariant derivative $ D_M=\nabla_M-\text{i}A_M $.
This is a classical system of coupled partial differential equations on the curved $4$-dimensional 
Anti-de Sitter spacetime \eqref{metric}. The Maxwell and Klein-Gordon 
equations, \eqref{Maxwell} and \eqref{KGE} respectively, are coupled through the electromagnetic current \eqref{current}.
This classical system of equations of motion together with the appropriate boundary conditions 
encode the quantum behavior of the dual field theory. 
We are interested in computing 
the full quantum evolution of the order parameter field $\psi=\langle \Psi \rangle$, where $\Psi$ is the 
operator dual to the complex scalar field $\Phi$ in the bulk. 
Finding the time evolution of the 
expectation value $\psi$ thus reduces to solving the classical equations of motion for $\Phi$.
To solve the equations of motion, we first need to impose boundary conditions for the gauge field $A_M$ as well as for 
the scalar field $\Phi$. 
The boundary condition for the temporal component $A_t$ of the gauge field at the conformal boundary $z=0$ fixes the chemical potential $\mu$ in the dual quantum theory, \ie,
\begin{align}
	A_t(t, \bm{x}, z)=\mu+\mathcal{O}(z)\,,
\end{align}
where $\mu$ is independent of the superfluid's coordinates $(t, \bm{x})$.

The only parameter of the holographic superfluid, uniquely characterizing it, is the dimensionless ratio $\mu/T\sim (\mu z_\text{h})$ (\cf\ \eqref{temp}).
One finds that the scalar field $\Phi$ forms a charge condensate above a critical value $(\mu z_\text{h})_\text{c}=4.06371$ (see \eg\ \cite{Herzog:2008he}) which sets the system in the superfluid phase. 
It is then straightforward to relate $\mu/T\sim (\mu z_\text{h})$ to the temperature ratio $T/\Tc$, where $\Tc$ denotes the critical temperature,
\begin{equation}\label{TvsMu}
\frac{T}{T_\text{c}}=\frac{(\mu z_\text{h})_\text{c}}{\mu z_\text{h}}\,.
\end{equation}
We fix our units by setting $z_\text{h}=1$. 
This also fixes the superfluid's temperature to $T=\frac{3}{4\pi}$ (see \eqref{temp}).
Therefore, $\mu$ is the only free parameter that controls the phase transition and hence the temperature ratio $T/T_\text{c}$ according to \eqref{TvsMu}.
Henceforth, we will characterize and discuss the superfluid in terms of its temperature ratio $T/T_\text{c}$ as this appears to be the more intuitive parameter. But we keep in mind that what we actually vary is the chemical potential.

According to the holographic dictionary, the scalar expectation value $\psi = \langle\Psi\rangle$ is 
identified as the next-to-leading order term in the near-boundary expansion of the scalar field $\Phi$,
\begin{align}\label{bcPhi}
	\Phi(t, \bm{x}, z)=\eta(t, \bm{x})z + \psi(t, \bm{x}) z^2 +\mathcal{O}(z^3)\,,
\end{align}
where $\eta(t, \bm{x})$ constitutes a source for the scalar field which has to be set 
to zero, \ie\ $\eta\equiv 0$, to not break the $U(1)$-symmetry explicitly. 
This condition can be expressed as
 \begin{align}
 	\partial_z\Phi(t, \bm{x},z)|_{z=0}=0\,.
 \end{align}
Similarly, all sources for the spatial gauge field components, which in the holographic interpretation 
correspond to an external superfluid flow, are switched off, \ie\ $A_x(t, \bm{x},z=0)=A_y(t, \bm{x},z=0)=0$.
At the black-hole horizon, $z=z_\text{h}$, the gauge field components must satisfy regularity 
conditions, given by $A_t(t, \bm{x},z_\text{h})=A_x(t, \bm{x},z_\text{h})=A_y(t, \bm{x},z_\text{h})=0$. 
Similarly, the scalar field $\Phi$ must behave regularly at the horizon which is 
intrinsically implemented by the choice of infalling Eddington--Finkelstein coordinates. 
More details on the equations of motion and the boundary conditions are given in appendix \ref{eom}.

The scalar order parameter $\psi$ is a complex field whose 
absolute value squared gives the density of the superfluid,
\begin{equation}
	n(t, \bm{x})=|\psi(t, \bm{x})|^2.
\end{equation}
We denote the homogeneous background density of the superfluid in thermal equilibrium, where no vortices are present, as $n_0$.
The phase $\varphi(t, \bm{x})=\text{arg}(\psi(t, \bm{x}))$ of the order parameter encodes the superfluid velocity
via $\bm{v}^\text{s}(\bm{x},t) = \vec{\nabla}\,\varphi(\bm{x},t)$, where $\vec{\nabla} = (\partial_x,\partial_y)$. 

In the probe approximation the black hole constitutes an infinitely sized static heat bath for the superfluid.
Due to its presence, the superfluid is equipped with an energy and momentum dissipation mechanism caused by modes falling into the black hole. 
In this sense, the black hole can loosely be interpreted as a normal component connected to the superfluid condensate in the sense of Tisza's two-fluid model \cite{Tisza1938TPiHII}.
In the case with back-reaction it has been shown that to non-dissipative 
order\footnote{This refers to the hydrodynamic expansion at the boundary with only non-dissipative terms included. 
	We stress that the near-boundary expansion \eqref{bcPhi} is not a hydrodynamic expansion in any sense. 
	Instead, it yields the order parameter field of the superfluid condensate that captures the physics at all scales.}
the holographic superfluid is indeed governed by a relativistic version of Tisza's two-fluid model \cite{Sonner:2010yx}.
It appears likely that a similar identification holds, at least approximately, also at dissipative order, \cf\ \cite{Sonner:2010yx}.
Also for the non-back-reacted system an analogy to the two-fluid model could be expected, but a rigorous
identification has not been established so far. Keeping this caveat in mind, it can be useful to view the results of
the following sections in light of the two-fluid picture. 

\subsection{Remarks on the Numerical Implementation}
\label{Subsec:NumImp}

We are interested in performing a real-time simulation of the dynamics of the holographic superfluid. 
The coupled system of equations of motion (see appendix \ref{eom}) cannot be solved analytically and hence must be dealt with numerically. 
The superfluid is described by the order parameter field $\psi$ 
which can be extracted from the solution of the equations of motion according to the near-boundary expansion \eqref{bcPhi}.
For $T/T_\text{c}<1$ the system is in the superfluid phase, \ie\ the global $U(1)$ symmetry is spontaneously broken. 
In section \ref{Sec:VortAntiVort} we will keep $T/T_\text{c}$ fixed at $T/T_\text{c}=0.68$ (corresponding to $\mu=6$)
and vary only the initial condition of our physical setup, that is the initial size of the vortex dipole. 
In section \ref{Sec:MuDependence}, on the other hand, we will keep the initial condition fixed and only vary the temperature, $T/T_\text{c}$.
All our numerical simulations are performed on a grid of $512\times 512$ points in the $(x, y)$-plane, with periodic boundary conditions imposed. 
For a given $T/T_\text{c}$ we choose the grid constant, $a$, such that an isolated vortex is resolved by $13$ grid points in diameter at $95\%$ of the background density $n_0$.
For $T/T_\text{c}=0.68$ this yields $a=1/3.5$. Identical grid constants are used in $x$- and $y$-direction 
such that an isolated vortex is resolved in a rotationally symmetric fashion. 
In section \ref{Sec:MuDependence}, where we discuss the dependence of our physical system on the temperature $T/T_\text{c}$, we give a table with all temperatures that we consider and the corresponding grid constants. 
For the holographic $z$-direction we use $32$ collocation points and accordingly a basis of $32$ Chebyshev polynomials. 
The system is propagated in time by an explicit and adaptive Runge-Kutta time-stepping scheme. 

In the following sections we often show vortex positions etc.\ at unit timesteps.
One unit timestep (also: unit of time) is composed of an adaptive number of much smaller timesteps $\tau\ll 1$
that we use in the solver for the differential equations. 
Further details on the numerical implementation are given in appendix \ref{App:Numerics}.

\section{Dynamics of Holographic Vortex Dipoles}
\label{Sec:VortAntiVort}

In this section we discuss the dynamics and the annihilation process of a single vortex--anti-vortex pair, or vortex dipole. 
Our main interest concerns the vortex trajectories and the vortices' local velocities and accelerations. 
We introduce a high-precision tracking algorithm for the position of the vortices that allows us to extract the trajectories in a quasi-continuous manner. 
It is this high precision that then enables us to accurately determine the velocity and acceleration fields of the vortices. 
Knowing the acceleration of the vortices we can in turn infer their (effective) interaction if we assume their (unknown) mass to be constant. 

Throughout this section the temperature of the superfluid will be fixed to $T/T_\text{c}=0.68$.

\subsection{Initial Conditions}
\label{Sec:InitialCond}

We study a system composed of only one vortex and one anti-vortex, separated by an initial distance of $u_0$ grid points.
This is the most basic non-trivial configuration and thus provides a clean environment for studying vortex dynamics. 
One possible source of numerical uncertainty originates from the periodicity of the grid and thus of the initial configuration.
To minimize it, we restrict ourselves to sufficiently small initial dipole sizes. In other words, we initialize our vortex dipole such 
that the `inner' distance, \ie\ the $u_0$ grid points in the initial condition, is much shorter than the `outer' 
distance, \ie\ the distance from vortex to anti-vortex obtained by crossing the boundary to the next periodic cell
of the lattice. (We always assume the dipole to be placed close to the center of a cell of the periodic lattice.)
With regard to this problem, we can access somewhat larger initial distances $u_0$ by placing the two vortices
on the diagonal of the quadratic grid. 
We perform all numerical simulations presented in this work on a grid of $512\times512$ points in the $(x,y)$-directions.
We have checked that our results are indeed independent of the grid size by performing one simulation on a $1024\times1024$
grid which gave identical results. We conclude that for all initial dipole sizes $u_0$ considered here the observed dynamics is not affected
by periodicity effects. When we align the vortices along the diagonal it is in general not possible to choose $u_0$ to be an integer.
For simplicity, we then quote the closest integer as the value of $u_0$. 

A vortex is a topological defect of the superfluid and has an intrinsic structure solely determined by the superfluid itself. 
The main property of a vortex is its phase winding $w$ around the vortex position at which the superfluid density vanishes. 
The quantized phase winding is given by a contour integral encircling the vortex position,
$w=\frac{1}{2\pi}\oint\,\text{d}\varphi$ for $\varphi=\text{arg}(\psi)$,
and is by definition positive for vortices and negative for anti-vortices. 

Vortices can be built into the system by multiplying the complex scalar expectation value $\psi$ by a localized 
phase, $\psi \to \psi \cdot e^{\text{i} w\phi_\text{v}}$, where $\phi_\text{v}$ is 
the appropriate phase angle in the $(x, y)$-plane. In our holographic system 
we multiply the complex scalar bulk field $\Phi$ by a localized 
phase $\Phi(t, \bm{x}, z) \to \Phi(t, \bm{x}, z)\cdot e^{\text{i} w\phi_\text{v}}$ at every $z$-slice along the holographic direction.
At the position of the vortex we in addition set the scalar field $\Phi$ to zero, again for all $z$-values. 
For the boundary superfluid this corresponds to a zero in the condensate density at the vortex position. 
As the system is evolved in time, the topological defect imprinted in this way fully builds up within a few unit timesteps
and becomes a vortex of characteristic size and shape.\footnote{The flow field for a vortex--anti-vortex pair
takes significantly longer to become fully developed, although a good approximation is reached within a few unit timesteps.
This will be discussed in section \ref{phasehealing} below.}

In figure \ref{Dens_Phase} we show one of the initial phase and density configurations (after
it has fully built up at $t=10$) used in this work. 
\begin{figure}[t] 
	\begin{center}
		\includegraphics{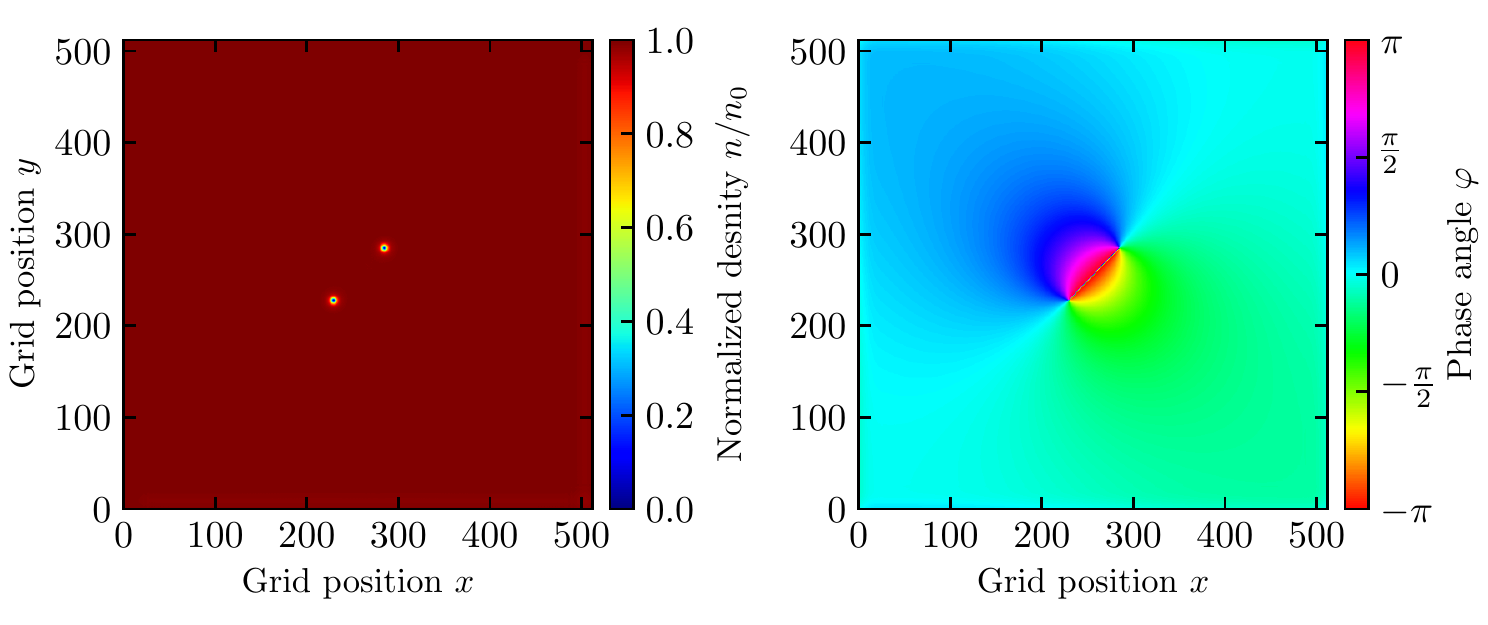}
		\caption{Initial condition for a typical vortex-dipole configuration used in this work. Left: Superfluid density with one 
			vortex and one anti-vortex aligned along the diagonal of the quadratic grid with initial separation of $u_0=80$ grid points. 
			The upper-right zero in the density is the vortex, the lower-left the anti-vortex.
			The superfluid density is scaled by $n_0$, the background condensate density without any defects.
			Right: The corresponding phase configuration of the system. 
			The color coding is chosen to be periodic to capture the nature of the periodic phase. 
			The snapshots of the density and 
			the phase are taken after 10 units of time, when the imprinted vortices are fully developed.
		\label{Dens_Phase}}
	\end{center}
\end{figure}
The vortex--anti-vortex pair is aligned along the diagonal
of the grid, initially separated by a distance $u_0$ of 80 grid points. The vortices actually start
moving immediately after they have been imprinted due to the phase configuration of the respective other vortex. 
When the vortex shape has fully built up the position of the vortices has therefore already changed and does no longer
exactly correspond to the initial distance. 
The plot on the left shows the superfluid density $n=|\psi|^2$ scaled to the background
condensate density $n_0$ (the equilibrium density of the superfluid without any topological defects). The plot on the right shows the 
phase of the order parameter field, $\varphi=\text{arg}(\psi)$. Both plots comprise the full grid. 
The initial conditions we study in this section are all similar to the one shown in figure \ref{Dens_Phase} and 
only differ in the chosen initial distance $u_0$ and the orientation of the dipole.

\subsection{Vortex Dynamics}
\label{Sec:Dynamics}

\begin{figure}[t]
	\centering
	\includegraphics[trim={0 0 0 1.1cm}]{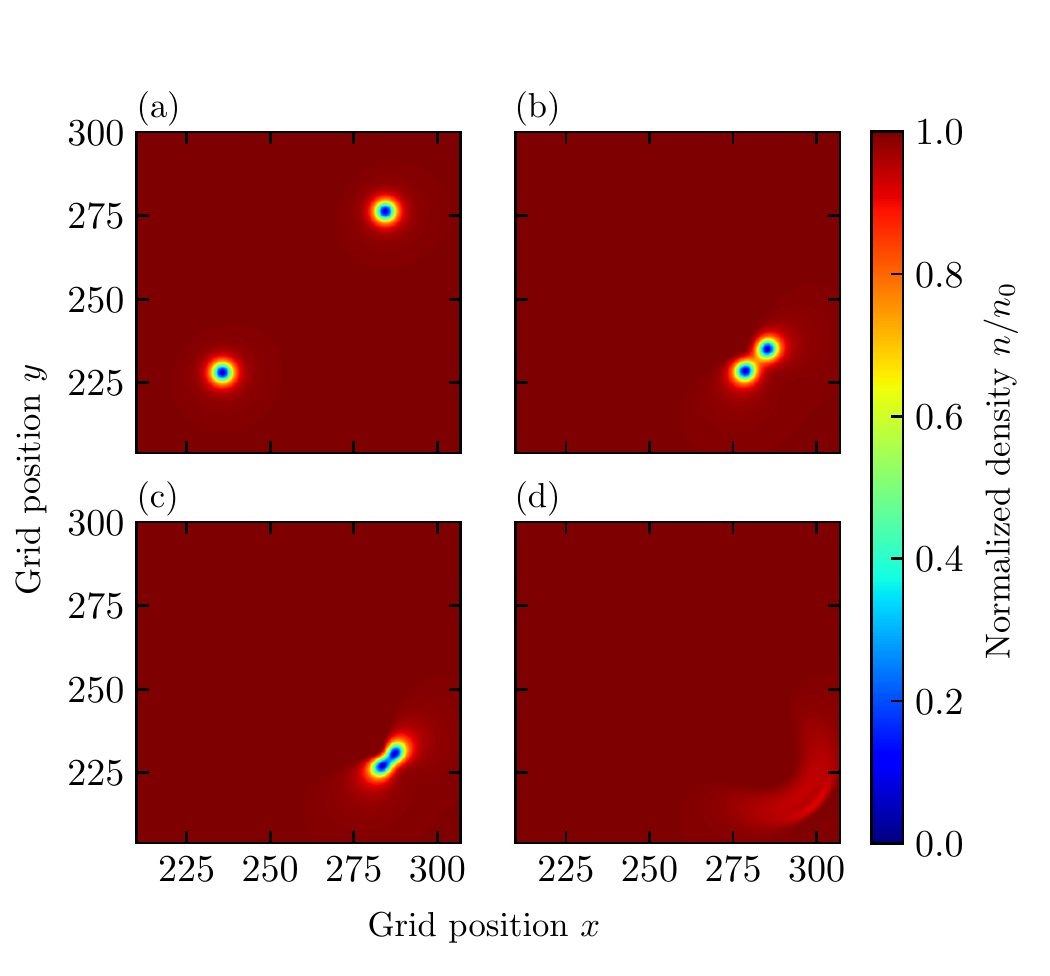}
	\caption{Snapshots of the superfluid density during the evolution of a dipole of initial size of $u_0=80$ grid points,
          taken at four different times (a)\;$t=400$, (b)\;$t=1530$, (c)\;$t=1546$, and (d)\;$t=1558$. 
		The initially circular vortices deform as they approach and finally annihilate each other. 
		At time $t=1558$ in plot (d) the vortices have already annihilated, producing a soliton-like excitation in the system. 
		This excitation is quickly damped to sound waves and eventually dissipated into the thermal bath of the system. 
	\label{Snapshots_4} }
\end{figure}

As the system evolves in time the vortices move on characteristic trajectories, approaching each other while simultaneously 
performing a motion in the perpendicular direction. 
We denote the motion along the dipole axis as longitudinal motion, and the one perpendicular to it as transverse motion.
In figure \ref{Snapshots_4} we show four snapshots of the superfluid density during the dipole evolution 
for an exemplary initial separation of $u_0=80$ grid points 
at different times, from upper left to lower right with proceeding time. 
It takes $t=1551$ unit timesteps for the vortices to annihilate for the chosen initial condition.

The first snapshot (a) is taken at time $t=400$, \ie\ early in the evolution when the vortices are far apart and still perfectly spherical. 
At this time the width of the vortices is much smaller than the vortex separation and the vortices can thus be interpreted as point-like defects in the superfluid density. 
We will come back to this in section \ref{hvisection} where we compare the trajectories to equations of motion in the point-vortex approximation. 
Snapshot (b) is taken at time $t=1530$, already shortly before the annihilation. 
At this stage, the superfluid depletions of the vortices start to deform non-elliptically and subsequently merge. 
This merging process begins only at very late time, here $\sim 25$ timesteps before the final annihilation, and proceeds very quickly. 
The transition from non-overlapping to overlapping density depletions is of course not sharp, but one can
still reasonably well estimate the time when it starts.  
The transition will be relevant for the precise tracking of the vortices, see section \ref{Sec:Tracking} below. 
In the following, we will refer to the last $25$ timesteps as late times.
In snapshot (c), four unit timesteps before the annihilation, the vortices have merged into a peanut-shaped
system with strongly overlapping density depletions. 
At this time, the dipole accelerates most strongly.
After the annihilation, we observe a soliton-like excitation propagating 
along the transverse direction of the vortex-dipole. This is shown in snapshot (d). The soliton-like excitation
transitions into sound waves. 
Each annihilation event of a vortex--anti-vortex system therefore releases energy 
into the system in the form of sound waves. 
The sound waves eventually dissipate completely, leaving the system in a homogeneous
equilibrium state. 

\subsection{Tracking Algorithm}
\label{Sec:Tracking}

So far, in all real-time simulations of holographic superfluids vortices were tracked by locating their
phase windings on the numerical grid, see \eg\ \cite{Adams:2012pj,Ewerz:2014tua,Lan:2018llf,Xia:2019eje,Du:2014lwa,Chesler:2014gya,Li:2019swh,Yang:2019ibe,Lan:2016cgl}.
The same holds for many studies of superfluids using the Gross--Pitaevskii equation (GPE)\footnote{The Gross--Pitaevskii equation \cite{Gross1961a,Pitaevskii1961a} or nonlinear Schr\"odinger equation is a mean-field description for the classical order parameter field $\psi$ of dilute ultra-cold Bose gases in the ground state, \ie\ Bose-Einstein condensates.}, see for instance \cite{Numasato2010,Karl:2016wko,Shanquan2020}. 
This procedure yields the vortex positions with plaquette resolution which is sufficient for many applications. 
We find, however, that to study and resolve the full dynamics of a vortex dipole a sub-grid-spacing resolution is required. 
Only then can velocities and accelerations be accurately computed. 
There exists a number of tracking methods in the literature, see \eg\ \cite{Nazarenko_2006,PhysRevE.78.026601,PhysRevE.86.055301,Zuccher_2012,Taylor_2014,rorai2014approach} for applications of different methods, each with different advantages and drawbacks. 
For the purpose of this work we propose a new tracking algorithm for vortices on a two-dimensional grid
and combine it with an existing Newton--Raphson (NR) tracking method, see for example \cite{GALANTAI200025}. 
This ensures that at each stage of the evolution the vortices are most accurately and numerically efficiently tracked. 

\subsubsection{Gaussian Fitting Routine}

In the following, we introduce and discuss a new tracking routine that has the advantage of being
numerically very efficient and applicable for most of the evolution of the vortex dipole.  
At a given time $t$ of the evolution we consider the condensate density $n(t, \bm{x})=|\psi(t, \bm{x})|^2$ in a subregion\footnote{The results are independent of the chosen subregion as long as it is large enough to fully enclose both, the vortex and the anti-vortex. In practice we choose a rectangle spanned by the vortex dipole plus $40$ grid points in the positive and negative $x$- and $y$-directions.} that encloses both vortices.
We then use a fitting routine that locates the minima of the density depletions, \ie\ the vortex positions, based on an interpolation 
between grid points. This permits a determination of vortex positions with sub-plaquette precision. 
As fitting function we use a linear combination of two two-dimensional upside-down Gaussians, 
\begin{align}
	G(x, y)=A&-B\exp\left\{-\frac{(x-x_{1})^2}{2\sigma_{x,1}^2}-\frac{(y-y_{1})^2}{2\sigma_{y,1}^2}\right\}\\
	&-C\exp\left\{-\frac{(x-x_{2})^2}{2\sigma_{x,2}^2}-\frac{(y-y_{2})^2}{2\sigma_{y,2}^2}\right\}\,,
\end{align}
where $(x_{1},\,y_{1})$ and $(x_{2},\,y_{2})$ are the $(x,\,y)$-positions of the vortex and anti-vortex, respectively. 
The positions are the fitting parameters we are mainly interested in. 
The widths $\sigma_{x/y,i/j}$ of the Gaussians in $x$- and $y$-direction are also separate fitting parameters to allow for non-spherical deformations of the vortices. 
Finally, $A, B$ and $C$ are additional parameters in the fit, accounting for the background density and a scaling of the Gaussians.
For the fitting routine we use a Levenberg--Marquardt least-squares fitting algorithm.

The reason that this fitting routine accurately tracks the vortex positions is the following. 
Essentially, the method relies on the fact that an extremum of a function of simple and approximately known shape can be found with high precision 
even when its values are known only on a much coarser discrete lattice. 
Vortices are characterized by their quantized phase windings which implies that the superfluid density vanishes at their exact location.
Consequently, to track the vortices one has to find these minima.
For spherical or elliptically deformed vortices the position of the minimum of the density depletions is fully encoded in the flanks.
Thus, one only has to accurately describe the flanks (in our method by Gaussians) and interpolate to the minimum.
The minimum of the Gaussian then coincides with the vortex location. 
A Gaussian turns out to be a very good functional approximation in the region of the flanks and can thus be 
used to determine the vortex positions.
We point out that in order to most accurately describe the flanks, the Gaussian fit should not be
forced to reach zero density since it does not capture the full vortex shape. 

Notably, any other function that describes the flanks of the vortices similarly well could also be used.
We have checked this explicitly and found very good agreement between the results. 
For instance, another choice could be the approximate shape \cite{Schakel2008,Pethick2006} of a vortex from
the Gross--Pitaevskii equation, given by 
\begin{align}\label{GPEVortex}
n(r) = \frac{n_0 r^2}{2 \xi^2  + r^2}\,,
\end{align}
where $n(r)$ is the 
density profile with $r$ denoting the radial distance from the vortex center,
$n_0$ the background density, and $\xi$ a free parameter determining the width of the vortex. 
Using this function in the fitting routine yields a discrepancy with the Gaussian fit of approximately $\pm 0.02$ grid points in the vortex positions. 
We estimate that the error of the vortex position resulting from the Gaussian fits is of the same order of magnitude. 
This can be reasoned by testing how well a GPE vortex (\cf\ \eqref{GPEVortex}) captures the position of a grid-discretized Gaussian (of typical parameters
found in the fits to our vortices) and vice versa. 
A simple analysis show that the discrepancies are also of the order of a $10^{-2}$ grid points.
Furthermore, a comparison with known tracking methods in the literature yields the same estimate.   
In particular, we compared the results with the NR method discussed below and found excellent agreement.
In all plots shown in this work the error in the vortex positions is much smaller than the size of the plot markers. 

\begin{figure}[t]
	\centering
	\includegraphics{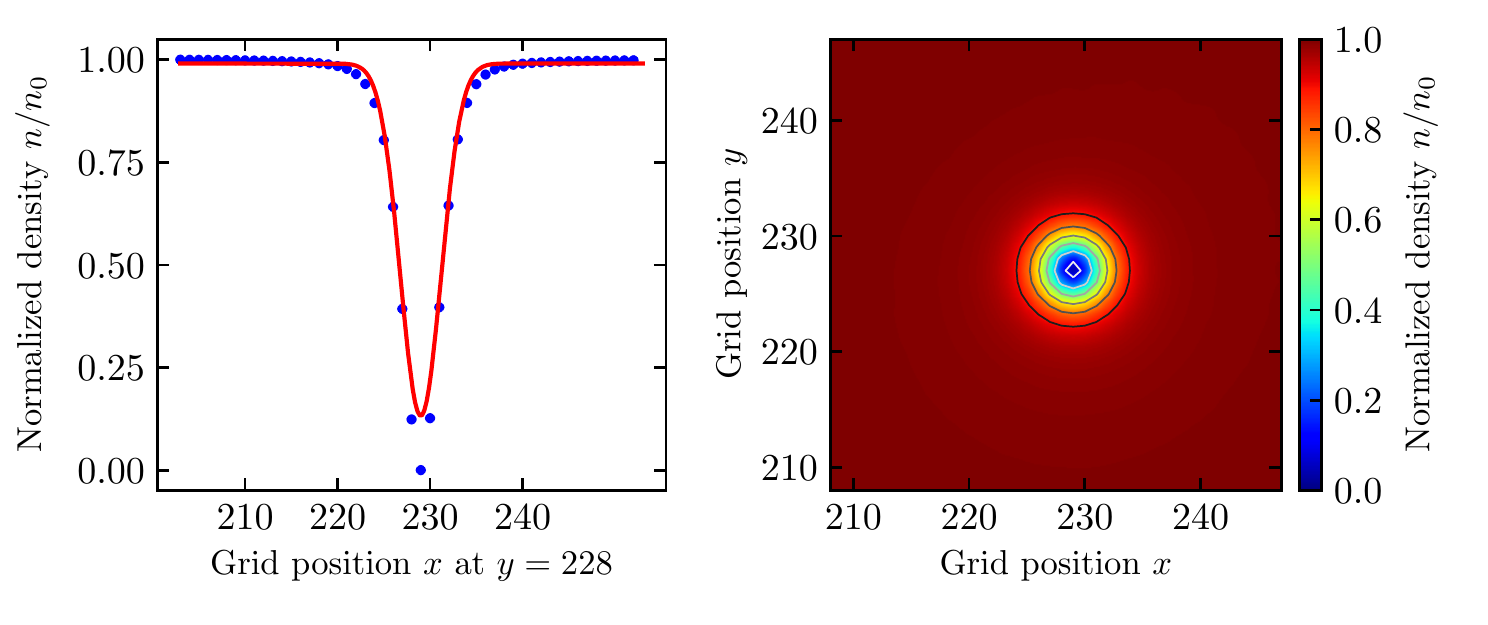}
	\caption{Graphical illustration of the Gaussian fitting routine for the localization of the vortices. 
		The plot on the left shows a cross section of the superfluid density (blue dots) along the $x$-grid for fixed $y$ and a 
		Gaussian fit of the data (solid red line). 
		The plot on the right shows the superfluid density in the vicinity of a vortex with contour 
		lines of the two-dimensional Gaussian fit in grayscale.
	\label{GaussFits}}
\end{figure}

Figure \ref{GaussFits} shows the cross section of a two-dimensional Gaussian fit to a vortex--anti-vortex system. 
Instead of showing the full dipole system we representatively focus on just a single vortex but stress that we in fact did a combined fit to both vortex and anti-vortex as described above.
The left plot in figure \ref{GaussFits} shows the cross section of the two-dimensional Gaussian fit to the 
superfluid density $n(x,y)$ along the $x$-direction for fixed $y$, with the $y$-position taken as the closest integer $y$-grid point to the position from the fit. 
Crucially, the Gaussian fit does not reach zero but captures the vortex flanks to high precision which is sufficient to find the position of the minimum. 
On the right in figure \ref{GaussFits} a contour plot of the same fitted vortex profile is shown. 
The contour lines of the fit are displayed in grayscale while for the condensate density $n$ we use the same color coding as in figure \ref{Dens_Phase}.

This Gaussian fitting algorithm can be used to track the vortices as soon as their shapes have fully developed after initialization, that is after $5$ unit timesteps. 
Only for the last $25$ unit timesteps, when the vortices merge and deform non-elliptically, this method breaks down and thus fails to track the vortices as the Gaussians can non longer accurately capture the vortex flanks. 
For this regime we instead rely on a well-known Newton--Raphson tracking algorithm which we discuss next. 

\subsubsection{Newton--Raphson Tracking Algorithm}

A reliable way to track the vortices also in the late-time regime when the density depletions of the vortices overlap is given by the NR method. 
As a standard algorithm for numerical root finding, the NR method is optimally suited for vortex tracking on two-dimensional grids, see for example \cite{PhysRevE.86.055301,PhysRevE.78.026601,Villois2016}.
In the following we give a brief overview of the method and discuss its advantages and drawbacks as compared to the Gaussian fitting routine. 
As discussed above, vortices are characterized by $\psi(\bm{x}_i)=0$, with $\bm{x}_i$ the position of the $i$th vortex, which is a direct consequence of the phase winding.
The NR method can therefore be used to solve $\psi(\bm{x}_i)=0$ directly, and no assumptions regarding the shape of the density depletion are needed. 
To ensure that a zero at $\bm{x}_i$ indeed signals a vortex one subsequently has to check that there is a corresponding phase winding at this position. 
To solve $\psi(\bm{x}_i)=0$ for $\bm{x}_i$ using the NR method, one performs a Taylor expansion of $\psi(\bm{x})$ around an initial guess $\bm{x}_i^g$ for the vortex position, 
\begin{align}\label{NRTaylor}
0=\psi(\bm{x}_i)=\psi(\bm{x}_i^g) + J(\bm{x}_i^g)(\bm{x}_i - \bm{x}_i^g) + \mathcal{O}\left[(\bm{x}_i - \bm{x}_i^g)^2\right]\,,
\end{align}
with the Jacobian matrix 
\begin{align}
J(\bm{x})=\begin{pmatrix}
	\operatorname{Re}[ \partial_x \psi(\bm{x})]   & \operatorname{Re}[ \partial_y \psi(\bm{x})] \\
	\operatorname{Im}[ \partial_x \psi(\bm{x})] &\operatorname{Im}[ \partial_y \psi(\bm{x})] \\
\end{pmatrix} \,.
\end{align}
If $J(\bm{x})$ is invertible, \ie\ has full rank, \eqref{NRTaylor} can be solved for $\bm{x}_i$, 
\begin{align}\label{NRiteration}
\bm{x}_i = \bm{x}_i^g - J^{-1}(\bm{x}_i^g)\,\psi(\bm{x}_i^g) + \mathcal{O}\left[(\bm{x}_i - \bm{x}_i^g)^2\right]\,.
\end{align}
This procedure can then be iterated such that $\bm{x}_i$ approaches the exact vortex position to sub-grid-spacing precision. 
The precision is mainly limited by the evaluation of the Jacobian at intermesh points. 
This is done by Fourier interpolating $\psi(\bm{x})$ which also guarantees that the periodicity of the order parameter field is retained throughout the iteration procedure. 
As initial guess for the algorithm we locate the phase winding of the vortices on the numerical grid. This plaquette precision is sufficient to ensure convergence of the NR algorithm. 
We stop the iteration when the condition $\psi(\bm{x_i}) < 10^{-10}n_0$ is satisfied.  

\begin{figure}[t]
	\begin{center}
		\includegraphics{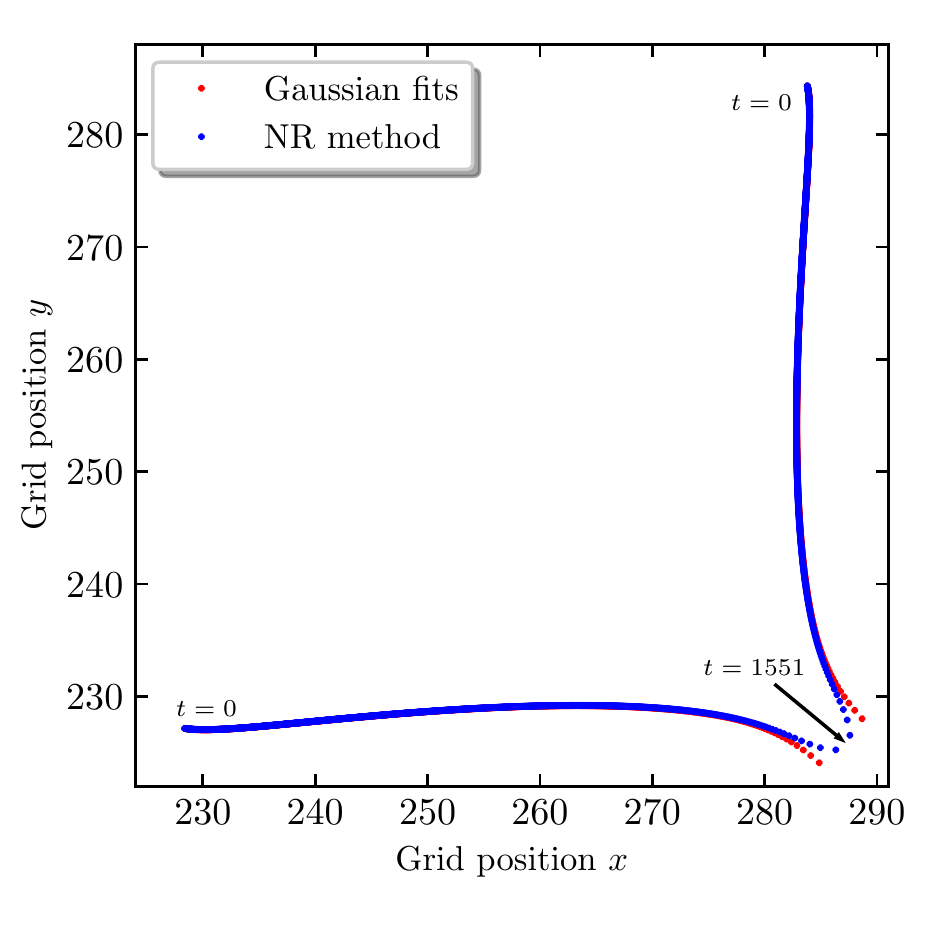}
		\caption{Comparison of the vortex trajectories obtained with the two different tracking algorithms,
                  for an initial separation of $u_0=80$ grid points. 
                  For most of the evolution, both methods give the same results with a discrepancy smaller than the marker size. 
			At late times the Newton--Raphson method is superior to the Gaussian fitting routine as it relies only on the zero in the superfluid density and not on the vortex flanks. Due to its numerical efficiency, the Gaussian fitting routine is superior for early and intermediate times, and hence for the larger part of the evolution.
		\label{GaussVsNR}}
	\end{center}
\end{figure}

The advantage of this algorithm as compared to the Gaussian fitting routine is that it directly tracks the zeros in the superfluid density.
The Gaussian fitting routine, on the other hand, determines the minima of the density depletions only via their flanks. 
That is very precise as long as the vortices deform only mildly and elliptically but fails as soon as this is no longer the case.
For the vortex dynamics under investigation this happens as part of the final annihilation process where the vortices deform strongly. 
For all earlier times, when the vortices are spherically symmetric or only slightly elliptically deformed, we checked that the two methods agree to a very high precision with a maximum absolute deviation of $\pm 0.02$ grid points. 
This error agrees with the estimate made above for the Gaussian fitting routine. 
The key advantages of the Gaussian fitting routine are its simplicity and its numerical efficiency which make it much faster than the NR method. 
Therefore, we use the Gaussian fitting algorithm to track the vortices for all vortex--anti-vortex separations $d(t)$ larger than $25$ grid points and the NR method for all smaller separations. 
As stated above, the two methods yield the same results for the first stage, \ie\ for $d(t)>25$.
We show a comparison of the trajectories obtained with the two methods in figure \ref{GaussVsNR}.
The difference in the vortex position at large separations is smaller than the marker size. 
For small separations, on the other hand, one clearly sees that the Gaussian fitting routine can no longer accurately track the vortices.
The trajectories obtained with this method cannot resolve the final annihilation process and have a systematic error,
as is evident from the almost parallel final pieces of the trajectories which would, if real, not lead to an annihilation of the vortices. 
The NR method, in contrast, provides the accurate vortex positions until immediately before the annihilation. 

\subsubsection{Sub-plaquette Vortex Tracking}

\begin{figure}[t]
	\centering
	\includegraphics{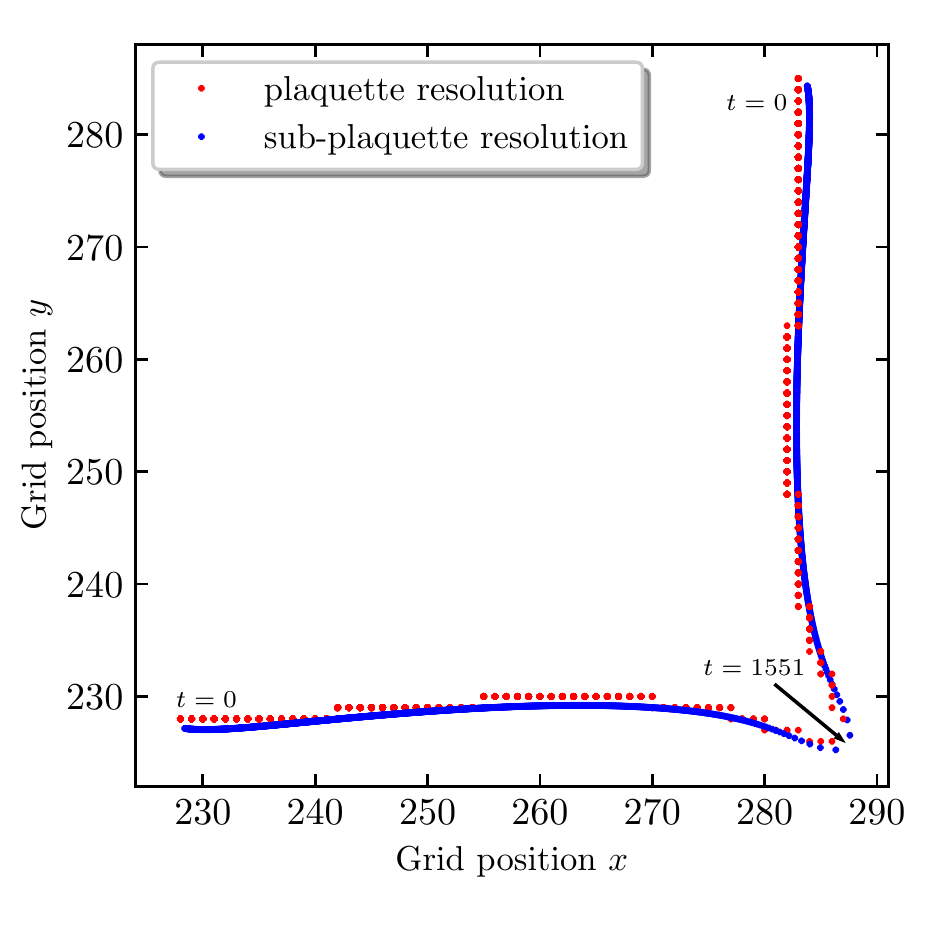}
	\caption{Trajectories of the vortex-dipole with an initial separation of $u_0=80$ grid points 
          computed by only localizing the phase windings of the superfluid phase (red dots)
          and with sub-plaquette resolution using our localization method (blue dots). 
          The phase-winding localization is only as precise as one grid point and the trajectory is therefore discrete or stair-like,
          while our localization method results in quasi-continuous trajectories. 
	\label{Trajectories} }
\end{figure}

Having discussed the tracking algorithms, we now focus on the gain of these methods as compared to the standard plaquette resolution obtained from locating the phase windings on the grid. 
The position found by locating the phase winding is inherently plagued with an uncertainty of one grid point 
as the vortex position is always assigned to integer grid points. 
Consequently the trajectories determined by this method take a stair-like shape, with 
the vortex position being allocated to one grid point for a certain number of timesteps before jumping to the next one. 
The tracking algorithms described above, on the other hand, allow to extract quasi-continuous trajectories (only limited by finite time-stepping resolution). 
In figure \ref{Trajectories} we compare the vortex trajectories determined by our (combined) tracking method to the trajectories 
determined by using only the phase windings to locate the vortices.
Clearly, the gain from the combined use of Gaussian fitting routine and NR method has a strong impact on the extracted trajectories.
We note that there is a systematic contribution to the deviation between the two methods coming from the fact that, when locating the vortices via their phase windings on a plaquette, we always attribute the position to the upper left corner of that plaquette. Any other choice would imply a similar systematic deviation that would be stronger in other parts of a given trajectory. 
Let us give an empirical estimate of the difference in precision between the two methods for our specific vortex-dipole system. 
At early and intermediate times (to be discussed in more detail below) it takes roughly $50$ unit timesteps for the vortex to move as far as one grid point. 
Locating the vortex solely by its phase winding would thus give the same grid point for all of these $50$ unit timesteps.
By using our fitting routine, on the other hand, we can clearly locate the vortex at $50$ distinct positions. 
Therefore, our method yields quasi-continuous trajectories for the vortices. 
As compared to the locations obtained from the phase windings this is a strong increase in precision.
Obviously, the gain in precision depends on the velocity of the vortices and, in relative terms, is smaller for fast vortex motion. 
We will see that the increase in precision due to sub-plaquette tracking
is essential for obtaining some of the main results in our analysis of the vortex dipole.  
For obtaining the data shown in figure \ref{Trajectories}, as well as for all other results presented in this work,
we choose $t=5$ as the starting time for our fitting procedure.\footnote{Note, however, that in figure \ref{Trajectories} the 
discrete trajectories of plaquette-resolution are shown starting at $t=0$.}
The trajectories are shown up to the point where the vortices annihilate.
Since the vortices accelerate strongly just before annihilation, we increase the frequency of tracking during the last 8 unit timesteps 
and determine the position 100 times per unit timestep here. 

\subsection{Trajectories of the Dipole System}
\label{DynamicsDipole}

\begin{figure}[t]
	\begin{center}
		\includegraphics{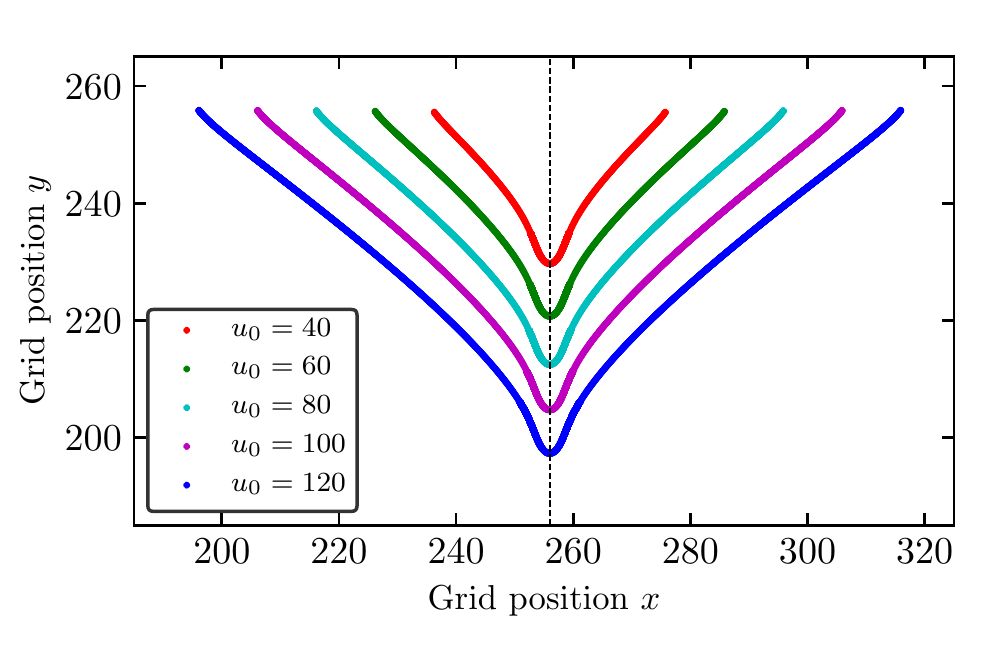}
		\caption{Trajectories of the vortex--anti-vortex dipole system with five different initial separations. 
                  The vortices are imprinted with initial separation $u=u_0$ along the horizontal of the quadratic grid.
                  (For details of the initialization see section \ref{Sec:InitialCond}.) 
			Here we show only the part of the grid containing the trajectories. 
			We observe that, after an initial phase healing process (see section \ref{phasehealing}), 
                        the trajectories become universal for all different initial separations.
		\label{VortPosBeide}}
	\end{center}
\end{figure}

Let us now investigate in detail the dynamics of a vortex--anti-vortex pair.
We first present and discuss the trajectories of the vortex 
dipole, here for initial separations along the horizontal of the quadratic grid. 
The results for five different initial separations are shown in figure \ref{VortPosBeide}.
Note that for conciseness the plot in figure \ref{VortPosBeide} does not 
show the full numerical grid but only the part containing the actual vortex trajectories. 
We observe that the trajectories of the vortices at intermediate and late times are independent of the initial separation and follow universal curves.
That is, the short-distance behavior is independent of the earlier evolution of the system. 
We checked this explicitly by shifting the trajectories such that the respective annihilations take place at the same point on the grid.
Figure \ref{VortPosBeide} shows the un-shifted vortex trajectories. 
We refer to all times from which on the trajectories are universal, but prior to the last $25$ unit timesteps, as intermediate times. 
We stress that the distinction between early, intermediate and late times in this sense is only approximate 
and does not split the evolution into three strictly distinguishable phases. 
The universality of the trajectories at intermediate and late times is a remarkable
feature of the system as it shows that the physics of a vortex--anti-vortex pair in the superfluid is
entirely governed by the size of the dipole and independent of its previous evolution. 
In subsection \ref{Velocities} we will discuss this point in more detail in the context of the vortex velocities and accelerations.

\subsubsection{Phase Healing}
\label{phasehealing}

In figure \ref{VortPosBeide} we observe a feature of the trajectories at early times that is, at first sight, very surprising
and gives rise to the non-universality of the trajectories in this regime. Both the vortex and the anti-vortex 
show a slight outward bending in their motion before they approach each other. The effect is discernible only when
the vortices are tracked with sub-plaquette resolution. This is obvious in figure \ref{Trajectories} where
the quasi-continuous trajectories (blue dots) exhibit a clear outward bending while the discrete trajectories (red dots) 
limited to plaquette resolution do not show any sign of outward bending.  

In figure \ref{VortPosBeide}, the outward bending is seen in the early-time evolution for all initial separations.
It turns out that the effect is not physical but is an artifact of the method used for imprinting the initial configuration
of vortices in the superfluid order parameter field. 
As discussed above, we initialize two single vortices by simply imprinting and linearly superimposing their individual phase
configurations on the entire grid. 
But this does not take into account that a proper vortex--anti-vortex pair is actually one coupled system.
Imprinting such a pair would require us to imprint the coupled phase configuration instead of the linear superposition of
two separate vortices. However, an analytic formula for the correct phase configuration of a proper vortex dipole is not
known. Imprinting the superposition of the phase fields of two separate vortices is therefore the best practical procedure.
Fortunately, during the numerical evolution of the system the phase configuration `heals itself' and transitions into that of a proper vortex dipole. 
This `phase healing' process is much slower than the build-up process of the vortex shapes, as can be expected because it
is necessarily a non-local process. After the phase healing the dynamics of the dipole becomes independent
of its initial configuration and is then solely determined by the distance between vortex and anti-vortex.

We refer to `intermediate times' of the evolution for times after the phase healing is completed, while we call
times during the phase healing process `early times'. Equivalently, intermediate times refer to the universal regime of the
evolution which is independent of the initial separation of the vortex--anti-vortex pair. We find that 
the phase healing process is mildly dependent on the initial condition, \ie\ the initial separation of the vortices, \cf\ figure \ref{VortPosBeide}.
However, the effect of the phase healing on the trajectories is rather subtle and fades out slowly. 
It is therefore difficult to determine the exact time of the transition from the phase-healing regime to the universal regime.
The analysis in section \ref{Velocities} will show that the vortex velocities permit a reasonably accurate estimate. 

For an initial separation of $u_0=80$ grid points we find that it takes about $\Delta t \sim 350$ unit timesteps
for the phase configuration to approach that of a proper vortex dipole. 
We can understand how the phase healing leads to a slightly outward pointing motion 
by comparing the phase configurations of two dipoles of identical size, with one of them being in its initial phase configuration
and the other one being in a phase configuration after the healing process. The latter can be obtained by evolving a larger dipole
through the phase healing. 
The difference between the two corresponding phase configurations gives the initially `missing configuration'
(as compared to the fully coupled dipole configuration), and we can derive the superfluid flow resulting from it. 

\begin{figure}[t]
	\begin{center}
		\includegraphics{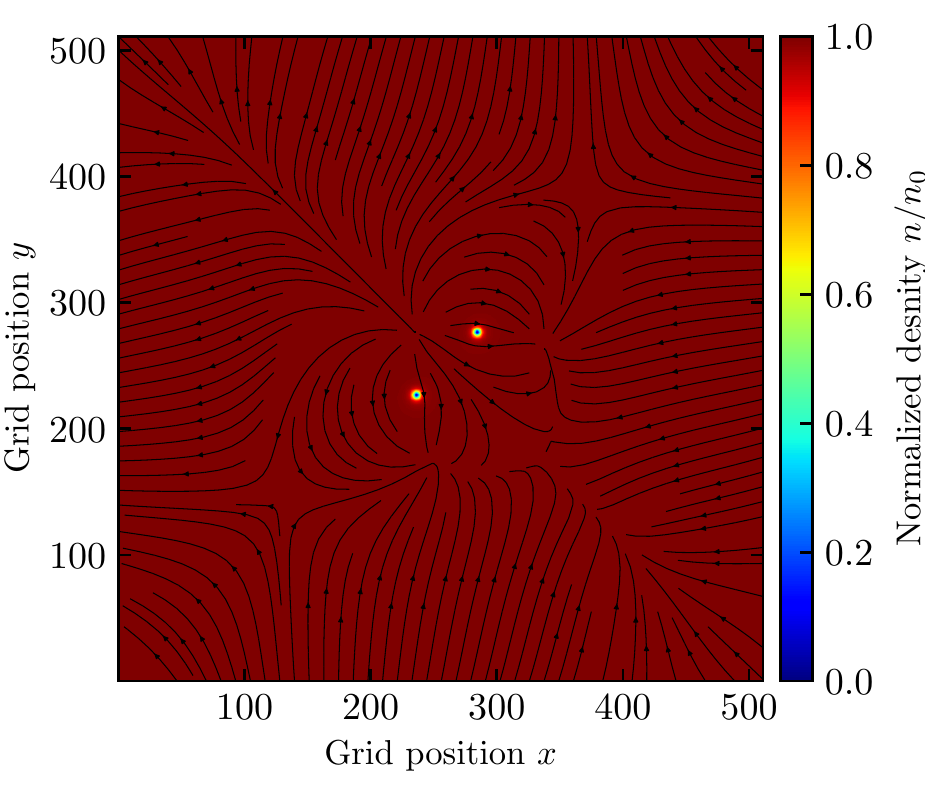}
		\caption{Density plot of a vortex configuration after the phase healing with remnant streamlines. 
                  The streamlines show the negative gradient of the difference of the phase field of a fully coupled (healed) dipole
                  and of the phase configuration used for initializing a vortex--anti-vortex system of the same separation. 
                  The phase of the evolved dipole, originally imprinted at a separation of $u_0=80$ grid points, is taken at $t=350$. 
		The outward pointing streamlines at the position of the vortices explain the initial 
		slight outward bending of the vortex trajectories (see figure \ref{VortPosBeide}). 
		\label{DenAndStream_Remnant}}
	\end{center}
\end{figure}

To visualize this effect, we initialize a vortex pair as a superposition of the individual phase fields at large distance
and propagate it in time for long enough so that its phase field has fully healed. From the corresponding
phase configuration we subtract the phase configuration we would have imprinted for a new pair of vortices
at this very separation. This gives us the missing phase configuration of the newly initialized dipole
as compared to a fully coupled dipole. Its negative gradient
is the superfluid flow that acts on the newly imprinted vortices at this separation due to their missing phase configuration. 
We show the streamlines of this flow field superimposed on the superfluid density of the dipole
configuration in figure \ref{DenAndStream_Remnant}.
Here we have chosen the dipole orientation along the diagonal of the grid so that
the vortex dipole motion is pointed towards the lower-right corner of the plot.
We can clearly see the outward pointing streamlines that lead to the observed outward bending of the trajectories. 
Furthermore, we observe streamlines pointing in the direction opposite to the vortex motion.
In subsection \ref{Velocities} we will see that this leads to an initial phase of deceleration in the vortex motion.
Understanding the artifacts of the initialization procedure allows us to separate them from actual dynamics
of the vortex dipole by concentrating on the universal regime of the evolution. 

It is worth pointing out that related effects due to coupled phase fields have been observed in
the context of kinks in scalar field theory \cite{Christov:2018ecz,Christov:2018wsa}. 

\subsubsection{Comparison to Hall--Vinen--Iordanskii Equations}
\label{hvisection}

There exist well-known equations of motion, first introduced by Hall, Vinen and Iordanskii (HVI) \cite{Hall1956a.PRSLA.238.204,Iordansky1964a.AnnPhys.29.335}, for the dynamics of a number of vortices in the point-vortex approximation. 
They can be derived on very general grounds in a hydrodynamical setup, taking into account the interactions of the vortices
with the normal and the condensate components of the fluid. 
The forces that drive the vortex dynamics are the Magnus force, orthogonal to the vortex motion, and drag forces. 
The equations of motion are derived by setting the sum of all forces in the closed system to zero and solving for the vortex velocities \cite{Ambegaokar1980a,Sonin1997a.PhysRevB.55.485}. 
Applied to our setup,  there is no velocity of the normal component of the fluid,
and each vortex only feels the velocity field created by the respective other vortex. 
The coupled differential equations for the two-dimensional trajectories $\bm{x}_i$, where $i \in \{ 1,2 \}$ labels the vortices, are then given by 
\begin{equation}\label{eomVortices}
\frac{\text{d}\bm{x}_i}{\text{d}t}=(1-C)\bm{v}^\text{s}_i -w _i C' \bm{e}_\perp\times \bm{v}^\text{s}_i \,.
\end{equation}
Here, $C$ and $C'$ are phenomenological friction coefficients characterizing the mutual friction between the vortices and the fluid,
$w_i$ denotes the winding number of the $i$th vortex, and $\bm{v}^\text{s}_i$ is the superfluid velocity induced by the respective other vortex.
We further denote by $\bm{e}_\perp$ a unit vector perpendicular to the $(x,y)$-plane in a right-handed coordinate system,
pointing in an auxiliary direction used only for a convenient notation of this equation and not to be confused with the holographic direction.

For a given set of initial conditions $\bm{x}_i(0)=\bm{x}_{0,i}$, equation \eqref{eomVortices} can straightforwardly be solved.
For one vortex $(w_i=1)$ and one anti-vortex $(w_i=-1)$ we give a derivation of these solutions in appendix \ref{VortexTrajCGLE}.
It turns out that the vortices move along straight lines and annihilate under a fixed angle, depending on the parameters of the theory. 

As we have seen, vortices in the holographic superfluid have a finite extension and in general cannot be considered as point-like defects. 
However, at large vortex--anti-vortex separation as compared to the vortex width, a point-vortex approximation is possible as far as
the effective interaction between the vortices is concerned. 
We can therefore compare the analytic results from the HVI equations to the vortex trajectories discussed above. 
Figure \ref{VortPosBeide} clearly shows that the vortices do not follow straight lines for most of the evolution. 
We note in particular that at early times, when the vortices are furthest apart, the motion is affected by the phase healing
and thus not expected to obey the equations of motion for a proper dipole. 
Furthermore, we find that after the phase healing the vortices quickly accelerate
and approach each other, again rendering the point particle approximation invalid. 
This is clearly visible in the vortex trajectories as the final annihilation process with overlapping density depletions is
preceded by a strong bending of the trajectories which sets in not long after the phase healing. 
Only in between these two regimes a possible straight segment of the trajectories may be found. 
It turns out that for most of the initial separations we study here, this segment, if it exists, is too short to be properly identified. 
For an initial separation of $u_0=120$, however, we indeed find that the vortices follow straight lines
for a short period of time after the phase healing and prior to the strong bending and ensuing annihilation. 
We present more details and a plot illustrating this behavior in appendix \ref{VortexTrajCGLE}.
The phase healing ultimately is an artifact of our method of preparing the initial state.
Thus we find that for sufficiently large separation the fully developed holographic dipole system indeed obeys
the HVI equations of motion.

Comparing in detail the evolution of a vortex dipole in the holographic superfluid,
in the Gross--Pitaevskii framework and in the HVI equations, 
using the tracking methods developed in this work, offers even the possibility to determine
the characteristic HVI friction coefficients of the holographic superfluid, see \cite{Wittmer:2020mnm}. 

\subsection{Size of the Dipole System}
\label{Dipole}

\begin{figure}[t]
	\begin{center}
		\includegraphics{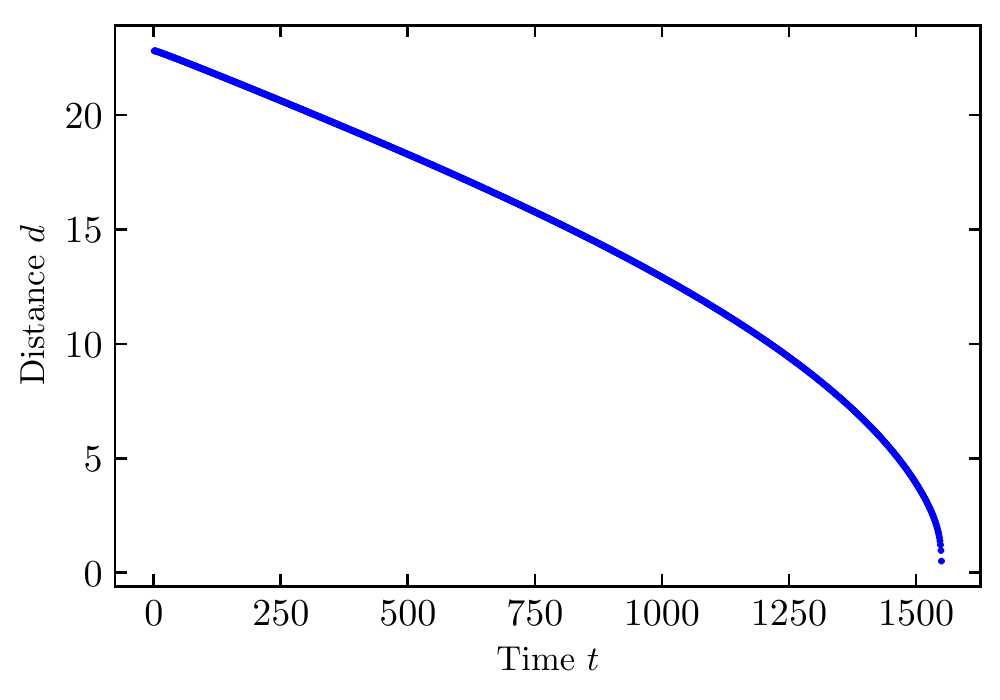}
		\caption{Size of the vortex dipole defined as distance between 
			the vortex and the anti-vortex as function of time, starting from 
			an initial separation of $u_0=80$ grid points along the diagonal of the grid.
                        The size in physical units $d(t)$ is obtained as $d(t)=u(t) a$, where in this example 
                        the grid constant is $a=1/3.5$. 
			\label{dist}}
	\end{center}
\end{figure}

Next, we study how the vortex--anti-vortex separation behaves as a function of time. 
The separation, or size of the vortex dipole, in units of grid points is defined as
\begin{equation}\label{distance}
	u(t)=|\bm{x}_1(t) - \bm{x}_2(t)|\,,
\end{equation}
where $\bm{x}_i(t)$ denotes the position of vortex $i \in \{1, 2\}$ on the numerical grid. 
In physical units, the distance is given by $d(t)=u(t)a$, where $a$ denotes the grid constant as discussed above.
(We recall that in this section we consider the temperature $T/T_\text{c}=0.68$ for which a grid constant of $a=1/3.5$ is chosen.) 
In figure \ref{dist} we show the physical distance as a function of time, again for an initial separation of $u_0=80$ grid points.
For all other initial separations discussed in this work $d(t)$ has the same functional behavior, as we discuss in
more detail in appendix \ref{DipoleSize}.
We expect the dipole size at early times to be affected by the phase healing process and thus not to reflect
the correct behavior of a proper vortex dipole. Therefore we focus on intermediate and late times in the following. 

There we find that $d(t)$ follows a power law for all initial separations.
Specifically, we perform a fit to our data with the ansatz 
\begin{align}\label{PowerLaw}
	d(t)=A(t_0 - t)^b\,,
\end{align}
where $t_0, A$ and $b$ are free fit parameters, and find that excellent agreement is achieved.
According to \eqref{PowerLaw} one would expect $t_0$
to be the time at which the vortices annihilate. However, during the final stages of the annihilation process, the
behavior of the vortices is dominated by the deforming and merging density depletions and is not adequately 
described by the notion of moving vortices in the sense of a formula like \eqref{PowerLaw}.
We therefore exclude the final 5 unit timesteps from the analysis and leave $t_0$ as a free parameter of the fit. 
For all initial separations we find for the exponent $b= 0.53 \pm 0.02$.
We present more details of the fit and further discussion of the results in appendix \ref{DipoleSize}.

We compare this result to the behavior expected from the HVI equations \eqref{eomVortices} for point vortices. 
As discussed in appendix \ref{VortexTrajCGLE}, the HVI equations predict
the vortex--anti-vortex separation $d(t)$ to follow a power law of the form \eqref{PowerLaw}
with $b=0.5$, see equation \eqref{dPowerLaw}.
This is very close to what we find numerically for the holographic vortex dipole.
However, we note that the holographic vortices in general cannot be considered as point-like,
and deviations from the square root behavior are therefore expected. 
Nevertheless, our results show that also for the case of finite-width vortices their spatial separation
obeys a universal power law after the initial phase healing.

The dynamics of a vortex dipole and its annihilation in the holographic superfluid were also
addressed in the recent study \cite{Lan:2018llf}, including in particular a calculation of the 
time dependence of the dipole size. In the following we comment on how our results compare to 
those of \cite{Lan:2018llf}. We first note that we study exactly the same system at 
the same temperature. The methods for the numerical simulation are similar, with Fourier decomposition
in the $(x, y)$-plane and a pseudo-spectral method with an expansion in Chebyshev polynomials
in the holographic $z$-direction. A grid of $241\times 241\times 25$ points in the $(x, y, z)$-directions is
used in \cite{Lan:2018llf} while we use a larger domain of $512 \times 512 \times 32$ points. 
A key difference between our work and \cite{Lan:2018llf} is that we use a high-resolution method for tracking 
vortices while the analysis of \cite{Lan:2018llf} relies on tracking by phase winding, thus providing only
plaquette resolution. 
The authors of \cite{Lan:2018llf} observe that $d(t)$ splits into two regimes each of which is described by a power law,
with a transition at a separation of twice the vortex radius (defined as the radius where $99\%$ of the background density are reached). 
At large separations the square root behavior predicted by the HVI equations is found,
while an exponent of $2/5$ is found at small separations. Velocities, accelerations and effective interaction strengths
of the vortices are then derived from these power laws, in contrast to our calculation in which velocities and accelerations
are computed directly. 
With improved precision in vortex tracking, we find different results for $d(t)$. For large distances, we find a similar power law, 
although with a slightly different exponent. For small distances, we do not observe a second, distinct power law.
We further observe that the point-vortex approximation breaks down at distances much larger than twice the vortex radius. 
The likely reason for the different results is the higher precision of our study, but also the larger numerical grid might play a role. 

\subsection{Velocities and Accelerations}
\label{Velocities}

The high-precision tracking of the vortices allows us to directly compute their velocities and accelerations. 
We can use the quasi-continuous trajectories determined above to compute derivatives via finite differences,
as we describe in appendix \ref{fdm}. Furthermore, we can then determine the dependence of
velocities and accelerations on the dipole size which is interesting with regard to the observed universality
of the dipole motion.

The results of this section again rely heavily on the improved tracking procedure, in particular for the slow motion
of vortices at early and intermediate times. Here, a vortex needs many unit timesteps to traverse a single grid spacing.
Accordingly, the velocity calculated from tracking with only plaquette resolution would be zero for most of the time,
except for occasional jumps when the next grid point is reached. From quasi-continuous trajectories, on the other
hand, we can compute quasi-continuous velocities and accelerations. 

\subsubsection{Vortex Velocities}

\begin{figure}[t]
	\begin{center}
		\begin{subfigure}[t]{\textwidth}
			\centering
			\includegraphics{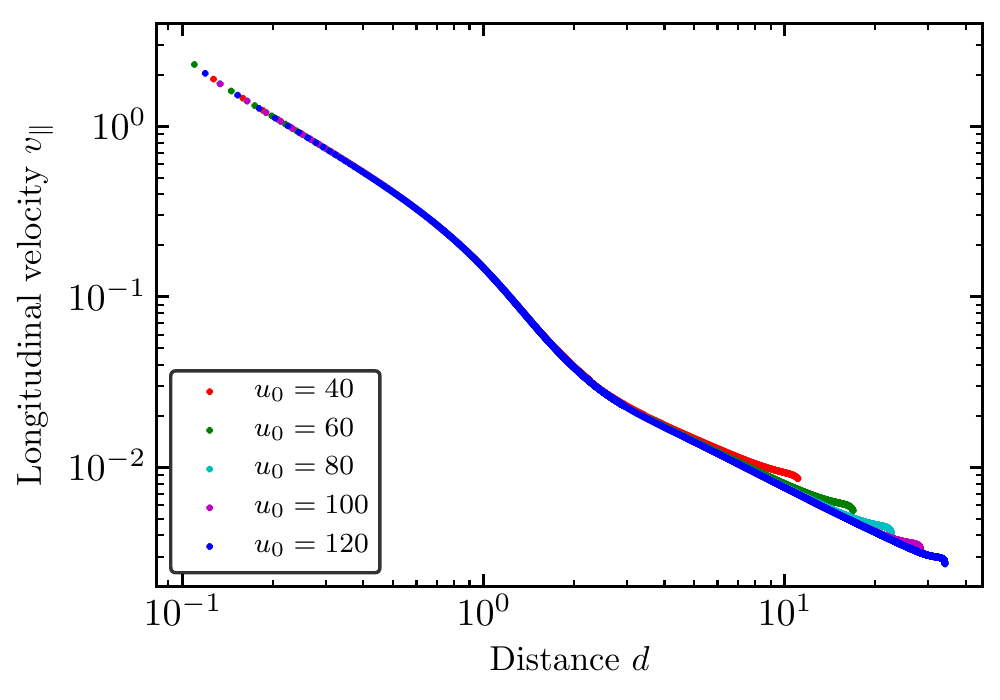}
		\end{subfigure}\vspace{0cm}
		\begin{subfigure}[t]{\textwidth}
			\centering
			\includegraphics{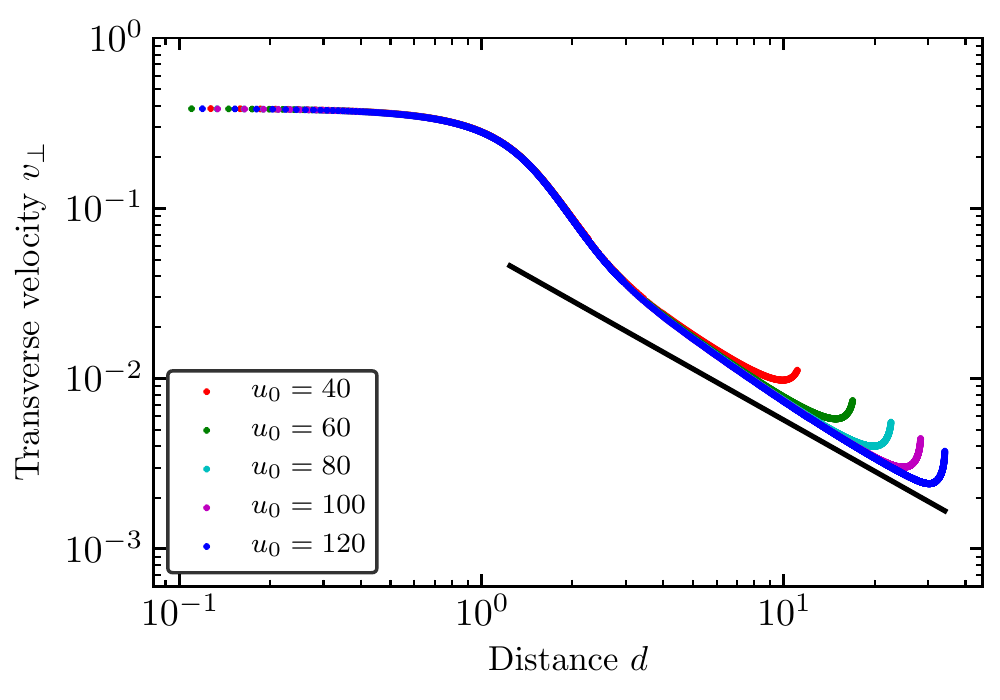}
		\end{subfigure}
		\caption{
			Longitudinal (upper plot) and transverse (lower plot) 
			components of the vortex velocities for five different initial separations on double-logarithmic scale. 
			Shown are the velocities for the anti-vortex. For the vortex the curves 
			are analogous except for an additional sign in the longitudinal component. 
			The velocities are given in physical units, \ie\ all lengths in grid points are multiplied by the grid constant $a=1/3.5$.
                        The curves show universality of the velocities at intermediate and late times.
                        The behavior of the velocities at early times can be attributed to the phase-healing process. 
			The straight line in the lower plot denotes an inverse-distance behavior for comparison. 
			Due to strong dissipative effects in the holographic system, deviations 
			from this behavior are expected.
                      \label{vels}}
	\end{center}
\end{figure}

In this section we consider physical distances (and velocities), \ie\ we use $d(t)=a u(t)$,
with $a$ the grid constant and $u(t)$ the dipole size in grid point. 
Consequently, velocities are expressed in units of the speed of light. 
To ease comparison with the trajectories discussed in the previous sections, we quote the
initial separation of the dipole $u_0$ in units of grid points. 
We recall that we can split the dipole motion into longitudinal and transverse parts.
Analogously, we denote the velocity component along the dipole axis as longitudinal velocity $v_\parallel$,
and the component perpendicular to the dipole axis as transverse velocity $v_\perp$. 
In figure \ref{vels} we show the two velocity components as function of the 
dipole size for all five different initial separations we considered before.
To wit, in these plots time proceeds from right to left as the dipole shrinks with increasing time. 
Figure \ref{vels} shows that after some time the velocities for all different initial separations approach a universal curve. 
This holds for both the longitudinal and the transverse component.
The deviations at early times are caused by the initial phase healing discussed above. 
At intermediate and late times, the velocities are universal and depend only on the size of the system,
confirming the universality observed in the trajectories in \ref{Trajectories}.
We also find that for most of the evolution the total velocity of the dipole system stays well below
the speed of light, such that the vortex motion can be considered non-relativistic.
This confirms the finding of \cite{Adams:2012pj} that the boundary dynamics of the holographic superfluid is
non-relativistic although the holographic model has an inherent Poincar\'{e} invariance. 

So far we have only considered the evolution of vortex dipoles of different initial separations. 
To test the universality of the vortex motion at intermediate and late times, we also performed simulations
with a different type of initial condition. For that we imprint the vortex and anti-vortex as before, but in addition
perturb the system with random noise\footnote{We perturb the system with noise by adding a random (real) phase $\theta (\bm{x})$ to the initial phase configuration. $\theta (\bm{x})$ is obtained by populating low-energy Fourier modes with (complex) random Gaussian noise and then transforming back to real space. For more details see \eg\ \cite{Ewerz:2014tua}.} at time $t=0$. Due to that, the vortices acquire a different initial velocity both
in absolute value and in direction. As a result, the trajectories at early times are very different from what we observed in figure \ref{VortPosBeide}.
At intermediate and late times, however, they can be shown to agree again after an appropriate rotation of the coordinate system (with an
angle depending on the initial motion due to the respective noise). 
Analogously, we find excellent agreement of the velocities and accelerations after the phase healing and after
the initial noise has dissipated. This again confirms the universal dynamics of the vortex-dipole 
in which the system loses information about its initial condition. 
In the following we will consider again the usual initial dipole configurations without noise. 

The phase healing at early times has a marked effect on the velocities. 
The transverse component of the velocity even decreases initially before it increases again.
We note that at this stage the velocity is exclusively induced by the initialized phase configuration
which, however, does not match the correct phase field of a coupled dipole. Therefore, the
velocities receive a contribution according to the streamlines shown in figure \ref{DenAndStream_Remnant}.
There, we indeed found a component pointing in the direction opposite to the center-of-mass motion of
the dipole. Once in the universal regime, on the other hand, the velocities exhibit the expected behavior
of a vortex dipole with a purely attractive force due to the strongly dissipative nature of the holographic superfluid
in the presence of vortices. 

In a superfluid condensate without dissipation, in contrast, vortices are known to exhibit so-called Helmholtz propagation
in which vortex and anti-vortex move at fixed distance on straight parallel lines. 
The velocity of the dipole is then purely dictated by the vortex--anti-vortex distance and follows
a simple inverse-distance law due to angular-momentum conservation in a closed system.
As our system is dissipative, Helmholtz propagation is suppressed and we expect deviations from the inverse-distance law. 
To illustrate this, we plot a $1/d$ curve in figure \ref{vels} for the transverse component of the velocity.

\subsubsection{Vortex Accelerations}

\begin{figure}[t]
	\begin{center}
		\begin{subfigure}[t]{\textwidth}
			\centering
			\includegraphics{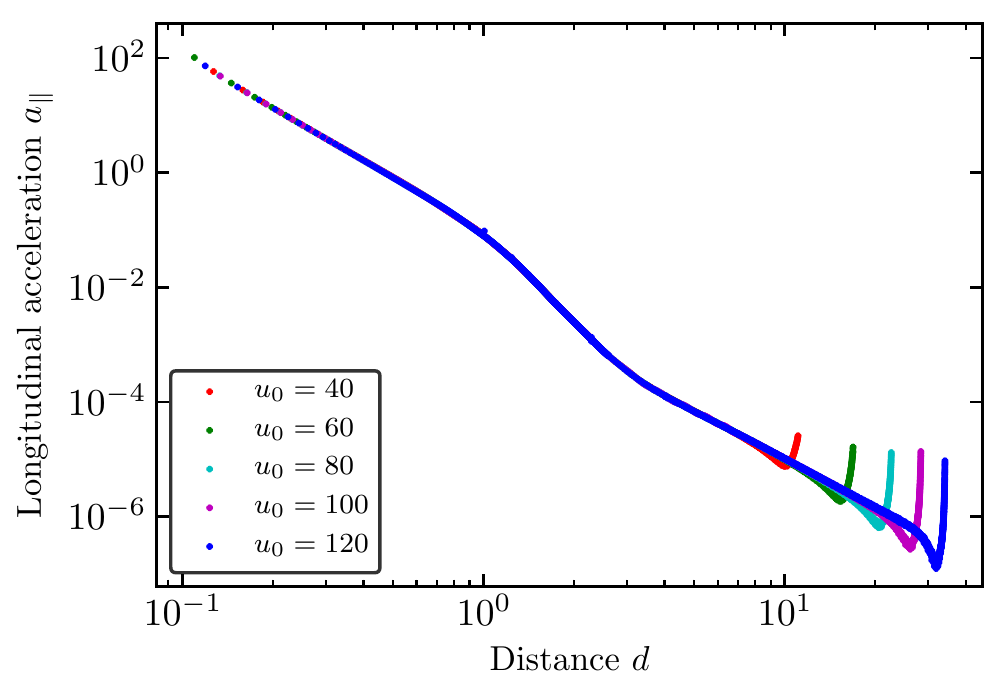}
		\end{subfigure}\vspace{0cm}
		\begin{subfigure}[t]{\textwidth}
			\centering
			\includegraphics{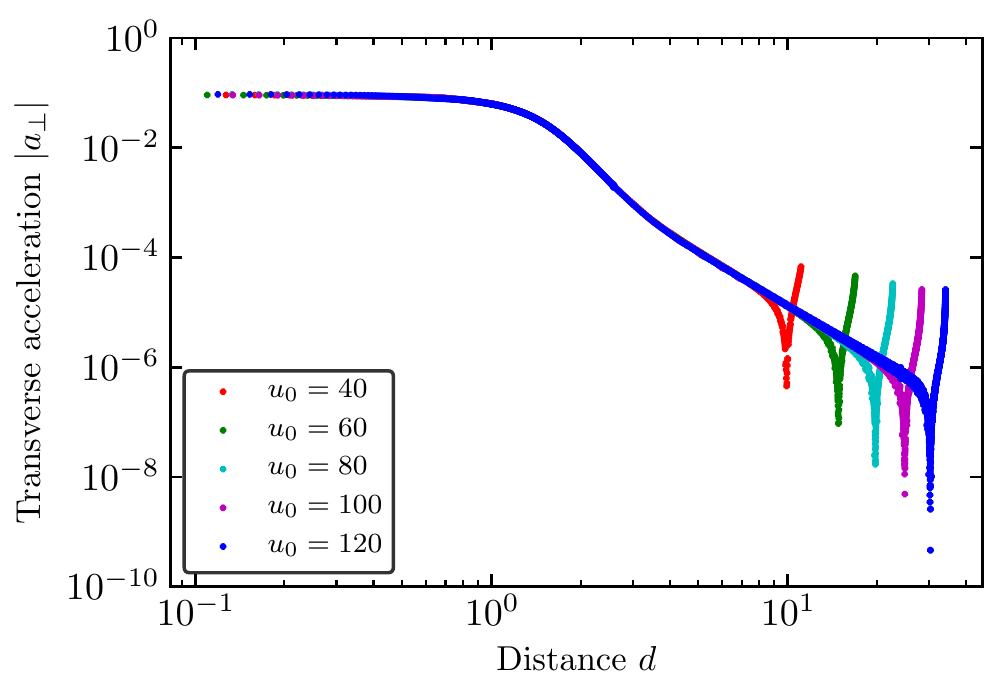}
		\end{subfigure}
		\caption{Modulus of the longitudinal (upper plot) and 
                  transverse (lower plot) components of the vortex and anti-vortex accelerations for five different initial conditions
                  on double-logarithmic scale. 
			All curves are displayed in physical units with a numerical grid constant of $a=1/3.5$.
			The kink in the transverse component and the initial behavior of the longitudinal 
			component can be attributed to the phase healing of the system.
		\label{accs}}
	\end{center}
\end{figure}

Although the velocity for given initial positions of the vortices
already contains the full information about the dipole motion, we still find it
instructive to show the corresponding acceleration, which we again 
split it into its longitudinal and transverse components. 
Figure \ref{accs} shows the longitudinal component $a_\parallel$ and the modulus of the transverse component $|a_\perp|$ of the acceleration
in physical units. 
The accelerations exhibit the same universal behavior at intermediate and late times after the phase healing. 
We note that the dips in $|a_\perp|$ are due to a change in sign of $a_\perp$. 

A comment is in order concerning the interpretation of vortex accelerations in terms of force fields.
Force fields usually require an interpretation in terms of test particles to be placed at a given position. 
However, such an interpretation is questionable for a quantized vortex as it always has a strong back-reaction on the fluid. 
Formally, one could of course compute acceleration vectors and interpret them in terms of a mutual force field acting
on the vortex and the anti-vortex. This would furthermore require the knowledge of an effective mass of a vortex. 
The latter problem has been addressed in the literature, mainly in the context of Gross--Pitaevskii theory, 
suggesting a definition for the mass of an isolated single vortex, see for instance \cite{Simula_2018}. 
Unfortunately, none of these models can be directly applied to our system and the effective vortex mass remains unknown. 
Nevertheless, we can infer from the accelerations in figure \ref{accs} that such an effective force depends only on the spatial separation
and is not a central force. 


\section{Vortex Dipole at Varying Superfluid Temperature}
\label{Sec:MuDependence}  

So far, we have considered vortex dipole dynamics only at a fixed temperature $T/\Tc=0.68$ of the holographic superfluid. 
In the following we compare the vortex velocities for a given initial separation while
varying $T/\Tc$. In general, we do not expect the probe approximation of the holographic superfluid to fully capture
all effects of the temperature. However, given that a fully back-reacted computation has not yet been performed, we
find it interesting to see how the dynamics changes with temperature. We keep in mind that the results might
be different in a fully back-reacted setting. 

\subsection{General Considerations and Initial Conditions}

\begin{figure}[t]
	\begin{center}
		\includegraphics{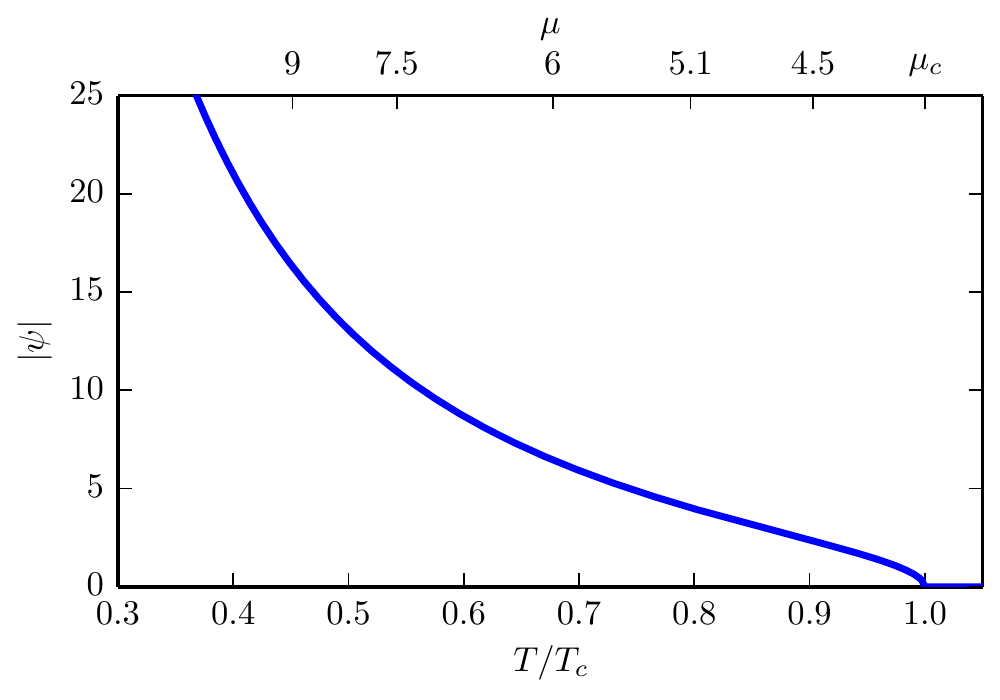}
		\caption{Order parameter as a function of temperature (lower abscissa) or 
			chemical potential (upper abscissa). As the temperature $T$ 
			falls below the critical temperature $T_\text{c}$ a condensate 
			forms, corresponding to a non-vanishing order parameter that 
			increases with decreasing temperature. 
			On the chemical potential axis we mark the values of $\mu$ used in this section.
		\label{cond_2d_T}}
	\end{center}
\end{figure}

We recall from section \ref{Sec:Model} that the holographic superfluid has only one dimensionless parameter $\mu/T$. 
Setting units by $z_\text{h}=1$ leaves only $\mu$ as free parameter which fixes the temperature $T/\Tc$.
In figure \ref{cond_2d_T} we show the modulus of the order parameter field in thermal equilibrium, which serves as our background
solution, as function of the temperature $T/T_\text{c}$ or the chemical potential $\mu$.
On the $\mu$-axis we mark those values that we use in this section.

Changing the temperature of the holographic superfluid implies a change in
the superfluid condensate density (see figure \ref{cond_2d_T}), a change in the time scales of the system,
and a change in the typical size of a vortex. In order to always resolve a vortex by the same number
of grid points we account for the changing vortex size by adjusting the grid constant for each fixed temperature.
This allows us to avoid any potential artifacts due to velocities and accelerations at different temperatures
being determined at different numerical resolution of the vortex structure. 
Again, all our numerical simulations are performed on a $512 \times 512$ grid in the $(x, y)$-direction.
We note that choosing different grid constants at different temperatures implies 
different physical sizes of the total system. 

We choose five different temperatures covering a broad range of the values in which the system is numerically accessible. 
Above and below our chosen temperatures the numerical solution turns out to be plagued by instabilities. 
One of the temperatures is the value chosen above, $T/T_\text{c}=0.68$ in the center of the accessible temperature range. 
In order to compare the dynamics of the system at different temperatures, we use dimensionless units for the quantities
in this section. 
To account for the dependence of the vortex size on the temperature $T/T_\text{c}$ we study the vortex 
velocities for a given $T/T_\text{c}$ as a function of the dimensionless distance $d(t)/\xi$, where $\xi$
is the vortex width as determined by fitting \eqref{GPEVortex} to the vortex profile for the given temperature.
Choosing instead the same initial separation in grid points would imply that the vortices are well separated for
some temperature while they would already start to overlap for another. 
We recall that for given temperature we choose the grid constant $a$ such that an isolated vortex is resolved
by $13$ grid points in diameter at $95\%$ of the background density $n_0$. 
For the initial separation we adapt $u_0$ such that $d_0/\xi$ (with $d_0=au_0$) it the same for the different temperatures.  

In table \ref{mu_params} we summarize our choices for the temperature, the corresponding chemical potentials,
and the resulting background densities $n_0$ and vortex width $\xi$. The initial separation $d_0$ is chosen
as $39.2 \,\xi$. In addition, we show the corresponding size $L$ of the computational domain in units of $\xi$. 
\begin{table}[t]
	\begin{center}
          \caption{Parameters used for different choices of temperature $T/T_\text{c}$ or chemical potential $\mu$.
            The vortex width $\xi$ and the initial dipole size $d_0$ have physical units. 
            Also given are the background condensate density $n_0$ and the size $L$ of the computational domain, here in
            units of $\xi$.
            \label{mu_params}}
		\begin{tabular}{cccccc}
			\toprule
			$T/T_\text{c}$&$\mu$ &$n_0$  &  $\xi$  &  $d_0/\xi$ & $L/\xi$ \\
			\midrule
			$0.90 $&$4.5$ & 	 $5.5 $ &  $1.16$  & $39.2$&$126.5$ \\
			$0.80$ &$5.1 $&	    $16.1$ & $0.85$ & $39.2$ & $171.1$\\
			$0.68 $& $6$ & 	  $41.7$ & $0.44$ &$39.2$  & $334.4$\\
			 $0.54 $& $7.5$ & $120.2$ &   $0.32$ &$39.2$ &$457.1$ \\	
			$0.45 $& $9 $   &$ 265.2$ &  $0.24$ & $39.2$ & $606.4$  \\
			\bottomrule
		\end{tabular}
		\label{tab:MuDynamics2D}
	\end{center}
\end{table}

\subsection{Velocities of the Vortex Dipole}

With these parameters we evolve the systems in time and track the vortices until they annihilate.
We find that the initial time it takes the vortices to fully build up their shapes after we imprint them 
is slightly dependent on the temperature but is always shorter or equal to $\Delta t=12$ unit timesteps 
which is well within the regime of the phase healing. As the velocities in this regime are anyway
affected by artifacts of the vortex initialization we can disregard any small differences in the
build-up process. We start the tracking for all chosen temperatures after the vortices have fully developed.
We point out that the tracking based on Gaussian fitting of the vortices works equally well for all temperatures
since the vortices are always resolved by the same number of grid points. 

\begin{figure}[t]
	\begin{center}
		\begin{subfigure}[t]{\textwidth}
			\centering
			\includegraphics{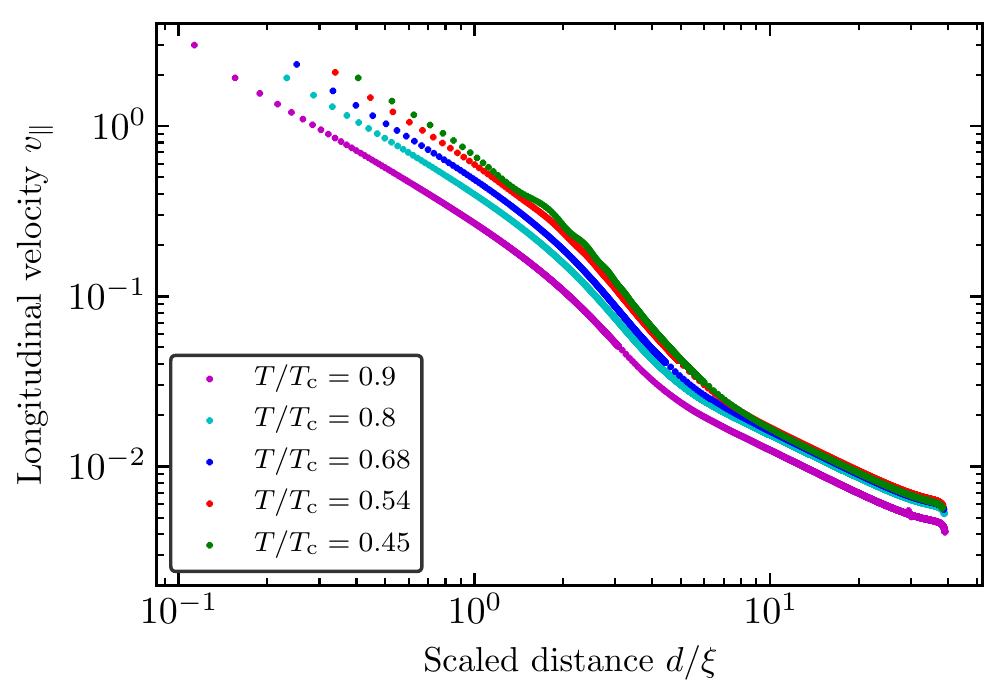}
		\end{subfigure}\vspace{0cm}
		\begin{subfigure}[t]{\textwidth}
			\centering
			\includegraphics{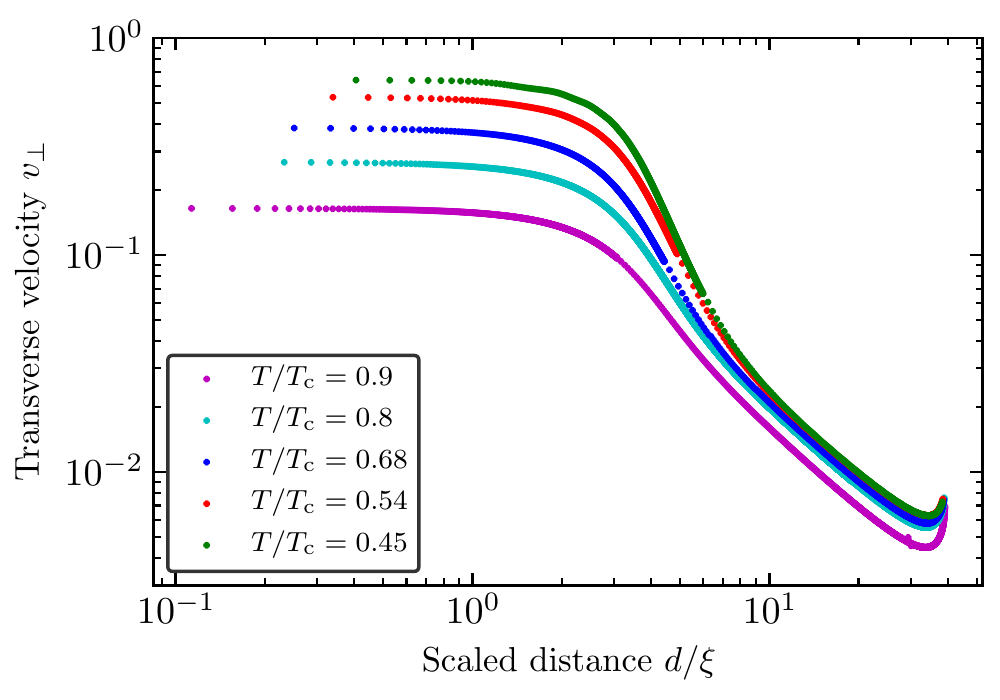}
		\end{subfigure}
		\caption{Longitudinal (upper plot) and transverse (lower plot) 
			components of the vortex velocities for five different temperatures. 
			The initial vortex separations $d_0$ are given in table \ref{mu_params}. 
                        Shown are the velocities for the anti-vortex. 
			For the vortex the curves are analogous except for an additional sign in the longitudinal component. 
			The effects due to the phase healing at early times (or large separations) are 
			stronger for higher temperatures.
                        \label{vels_mu}}
	\end{center}
\end{figure}

The longitudinal and transverse velocities, defined as before and again computed via finite differences from the tracked trajectories,
are shown in figure \ref{vels_mu}. 
The figure shows that the effect of the phase healing is present for all temperatures, but more pronounced at higher temperatures. 
We further see that after the phase healing process the velocities exhibit a clear ordering according with temperature. 
Both transverse and longitudinal velocities decrease with increasing temperature.

An interpretation of the ordering of the velocities with temperature in terms of dissipation, for example
along the lines of Tisza's two-fluid model, would appear speculative as
the results are obtained in the probe-approximation and might be affected when back-reaction is included. 


\section{Summary and Outlook}
\label{Sec:Summary}

We have studied the dynamics of a single vortex--anti-vortex pair in a holographic superfluid in numerical real-time simulations. 
A new tracking method has been introduced which allows us to locate the vortices with strongly improved precision
as compared to the conventional plaquette-based technique of locating the phase windings in the condensate density. 
The new tracking method relies on locating the minimum of the density depletion of the vortex rather than only its phase winding.
Using a fit with Gaussians, supplemented by a Newton-Raphson method at small vortex separations, we can determine the vortex
positions with sub-plaquette resolution in a numerically efficient manner. Fitting both vortices simultaneously with
a linear combination of two two-dimensional Gaussians even allows us to capture the mutual deformation at small distances,
thus eliminating a further source of error in the vortex positions. Our tracking method can be applied throughout the evolution
of the vortex dipole. With this method we have, for the first time for the holographic superfluid, computed quasi-continuous
trajectories of the vortices during their approach and eventual annihilation. In addition, we use large numerical grids of
$512 \times 512$ points in the $(x, y)$-directions such that finite-size (or periodicity) effects are efficiently suppressed, as
we have checked on a $1024 \times 1024$ grid. 

The increased precision makes it possible to study many aspects of the dynamics of the quantized vortices
which are not accessible with plaquette-resolution.
In particular, we are able to derive directly the velocities and accelerations of the vortices during the whole evolution of the dipole.
This allows us to improve earlier results of the pioneering study \cite{Lan:2018llf} in which the velocities were derived from
plaquette-resolution trajectories. We partially confirm the results found there for the power law for the velocity during the
dipole evolution. We find a very similar though not exactly equal exponent for large vortex separations, with our result indicating a
slight deviation from the behavior expected from the Hall--Vinen--Iordanskii equations for point-like vortices. 
However, we do not confirm the finding of a second power law for small distances reported in \cite{Lan:2018llf}. 

A key result of our study is the observation of a universal behavior of the vortex dipole in the holographic superfluid. 
Specifically, we find that its
trajectory and velocities are universal independent of the initial separation. Hence their interaction in the superfluid
is only determined by their separation and does not depend on the previous motion. This finding holds for all fixed
temperatures of the superfluid condensate. We point out that we observe in our simulations some 
deviations from this universal behavior at early times, visible as a slight initial outward bending of the trajectories.
But we have shown that these can be fully attributed to artifacts of the numerical initialization of the phase field of
the vortices. Such artifacts occur not only in the holographic superfluid but also in other numerical simulations of superfluids like for
example in Gross--Pitaevskii dynamics. It might be interesting to study this problem in more detail so that one could
potentially identify better methods for initializing ensembles of vortices in superfluid simulations. 
We note that our finding of universality in the vortex dynamics could again only be obtained due to the improved
precision of the vortex tracking. 

We have investigated how the dynamics of the vortex dipole depends on the temperature of the superfluid.
Our results indicate an overall decrease in the vortex velocities with increasing temperature if
we compare the system for suitably rescaled dimensionless quantities. However, the probe approximation
for the holographic superfluid might not be well suited to obtain a full picture of the temperature dependence.
It would therefore be interesting to study this question in a fully back-reacted system. 

The vortex--anti-vortex system studied here is one of the most basic non-trivial examples of topological
defects in a superfluid, and in some sense a basic building block for understanding more complicated
ensembles of vortices. Such systems of vortices and anti-vortices, as well as solitons, have
been widely studied for the holographic superfluid. We expect that the improved method for vortex tracking
can be very advantageous also for improving our understanding of such systems. It would for example be
interesting to see whether the universality of the dynamics of a vortex dipole has an equivalent for systems of
many vortices. 

The tracking method presented here can also be used for simulations of vortices in Gross--Pitaevskii dynamics.
An interesting result of such a study is the quantitative determination of the strong dissipation and of the friction
coefficients of vortices in the holographic superfluid \cite{Wittmer:2020mnm}. 
This result has been obtained by a high-precision comparison of vortex trajectories
in the holographic superfluid with those of Gross--Pitaevskii dynamics and of the Hall--Vinen--Iordanskii equations. 
Finally, we expect that our tracking method will also be useful in the study of vortex dynamics in
three-dimensional superfluids, including holographic superfluids. 

\acknowledgments
We would like to thank Thomas Gasenzer, Markus Karl, Christian-Marcel Schmied and Ulrich Uwer for helpful discussions, and Panayotis Kevrekidis for drawing our attention to \cite{Christov:2018ecz,Christov:2018wsa}. 
The work of P.\,W.\ was supported by the Studien\-stiftung des Deutschen Volkes e.\,V. 


\appendix
\section{Equations of Motion}
\label{eom}

Here we provide further details regarding the holographic superfluid discussed in section \ref{Sec:Model}.
In particular, we give details on the bulk equations of motion of the gravitational system dual to the superfluid. 
We recall that the action is given by 
\begin{align}
S=\frac{1}{2\kappa}\int\,\text{d}^4x\,\sqrt{-g}\left(  \mathcal{R}-2\Lambda+\frac{1}{q^2}\mathcal{L_\text{matter}}\right)\,,
\end{align}
where $\mathcal{L_\text{matter}}$ is the matter Lagrangian,
\begin{align}
\mathcal{L_\text{matter}}=-\frac{1}{4}F_{MN}F^{MN}-|(\nabla_M-\text{i} A_M)\Phi|^2 -m^2|\Phi|^2\,.
\end{align}
We work in the probe limit and thus neglect the back-reaction of the matter part on the gravity background. 
The resulting equations of motion are Maxwell's equation and a Klein--Gordon equation for the scalar field,
\begin{align}
\nabla_MF^{MN}&=J^N\,,\label{App:Maxwell}\\ 
\left(-D^2+m^2\right)\Phi&=0\label{App:KGE}\,,
\end{align}
with the current 
\begin{align}
J^N&=\text{i}\left(\Phi^\ast D^N\Phi-\Phi\left(D^N\Phi\right)^\ast\right)\,.
\end{align}
The remaining gauge freedom is fixed by choosing the gauge $A_z=0$. 
In our units with $L_\text{AdS}=1$ we fix the squared mass to $m^2=-2$.
We also fix the position of the black-hole horizon to $z_\text{h}=1$ which leaves us
with only one free parameter of the theory, the chemical potential $\mu$.
This can be tuned such that the system is in the superfluid phase. 

To compute the dynamics of the holographic superfluid we first have to calculate 
the static and homogeneous background solution. 
Subsequently we imprint the vortices into this background solution and evolve the system in time.
The static background solution is independent of time and of the spatial coordinates $x$ and $y$ of the superfluid 
but depends non-trivially on the holographic coordinate $z$. 
Plugging the AdS-Schwarzschild metric \eqref{metric} into the Maxwell equations \eqref{App:Maxwell} we obtain 
\begin{align}
0 &= z^2 A_t'' + 2\Im(\Phi'\Phi^*) \,,\\
0 &= z^2(h A_i''+h'A_i') - 2\lvert\Phi\rvert^2 A_i \,,\\
0 &= 2A_t\lvert\Phi\rvert^2 - \i h\left(\Phi^*\Phi'-\Phi^*{}'\Phi\right) \,, \label{eq:4Maxwell}
\end{align}
where a prime denotes the derivative with respect to the holographic coordinate $z$ and the index $i$ runs over the spatial coordinates $x$ and $y$.
The fourth Maxwell equation \eqref{eq:4Maxwell} describes the dynamics of $A_z$ which, by gauge freedom, we choose to vanish. 
Therefore, equation \eqref{eq:4Maxwell} remains as a constraint on the chosen axial gauge. 
For the scalar field we find from \eqref{App:KGE}
\begin{equation}
0 = z^2 h \Phi'' - z \left(-2\i z A_t + 2h -z h'\right)\Phi'
- \left(2\i z A_t - \i z^2 A_t' + z^2 \vec{A}^2 +m^2 \right)\Phi\,,
\end{equation}
with $\vec{A} = (A_x, A_y)$.
These equations are ordinary differential equations of second order in $z$. 
We already discussed the boundary conditions for the fields in section \ref{Sec:Model} but repeat them here for completeness.
The chemical potential is given by the boundary value of the temporal component of the gauge field, \ie\
\begin{align}\label{A0boundary}
A_t(z)=\mu+\mathcal{O}(z)\,.
\end{align}
At the black-hole horizon, $z=z_\text{h}=1$, $A_t$ has to be regular, $A_t(z=z_\text{h})=0$.
For the spatial components of the gauge field the boundary conditions at the conformal boundary $z=0$ of the AdS spacetime are given by
\begin{align}\label{Axyboundary}
A_x(z=0)=A_x(z=0)=0\,,
\end{align}
while at the black-hole horizon
\begin{align}
A_x(z=z_\text{h})=A_x(z=z_\text{h})=0\,.
\end{align}
As discussed in section \ref{Sec:Model}, the scalar field has to be regular at the black-hole horizon, which is 
ensured by the choice of infalling Eddington-Finkelstein coordinates.
At the conformal boundary the scalar field $\Phi$ has to satisfy
\begin{equation}\label{Phiboundary}
	\partial_z\Phi(z)\rvert_{z=0}=0\,,
\end{equation}
which sets the source of the scalar operator in the boundary theory to zero such that the $U(1)$ symmetry is not explicitly broken.

After imprinting vortices into the background solution, the system is no longer homogeneous and we have to solve the 
full set of equations of motion to obtain the real-time dynamics of the superfluid. 
For the numerical implementation it turns out to be convenient \cite{Adams:2012pj} to derive the full set of 
equations of motion in terms of the rescaled scalar field $\tilde{\Phi}=\Phi/z$ and to use the `lightcone derivative',
\begin{equation}\label{lightconeder}
\nabla_+X = \del_tX - \frac{h(z)}{2}\del_zX\qquad\text{for}\qquad X = A_x, A_y, \Phit\,,
\end{equation}
where $\vec{\nabla} = (\partial_x,\partial_y)$.
The full system of equations of motion is then given by
\begin{align}
\partial_z^2A_t &= \partial_z\vec{\nabla}\cdot\vec{A} - 2\Im (\Phit^{\ast}\partial_z\Phit) \,,%
\label{eq:1}\\[.6em]
\partial_z\nabla_+A_x &= \frac{1}{2}\left(\partial_y^2 A_x%
+\partial_x(\partial_zA_t-\partial_yA_y)\right)%
- \lvert\Phit\rvert^2 A_x + \Im (\Phit^{\ast}\partial_x\Phit) \,,
\label{eq:2}\\[.6em]
\partial_z\nabla_+A_y &= \frac{1}{2}\left(\partial_x^2A_y%
+ \partial_y(\partial_zA_t-\partial_xA_x)\right)%
- \lvert\Phit\rvert^2 A_y + \Im (\Phit^{\ast}\partial_y\Phit) \,,
\label{eq:3}\\[.6em]
\partial_z\nabla_+\Phit &= \frac{1}{2}\vec{\nabla}^2\Phit %
-\i\vec{A}\cdot\vec{\nabla}\Phit + \i A_t\partial_z\Phit %
-\frac{\i}{2}\left(\vec{\nabla}\cdot\vec{A} - \partial_zA_t \right)%
\Phit-\frac{1}{2}\left(z + \vec{A}^2\right)\Phit \,.
\label{eq:4}
\end{align}
The dynamical equation for $A_z$ is again used as a constraint equation and is not independent of equations \eqref{eq:1}-\eqref{eq:4}.

\section{Numerical Implementation}
\label{App:Numerics}

Here we discuss the numerical implementation of our solver for the system 
of differential equations given in appendix \ref{eom}. 

For the holographic $z$-direction we employ a collocation method with $32$ Gauß–Lobatto grid points and expand the 
fields in a basis of $32$ Chebychev polynomials.
After setting $z_\text{h}=1$, we have to use the rescaled coordinate $\tilde z\in [-1,1]$, defined implicitly via
\begin{equation}
z=\frac{1}{2}(\tilde z +1)\,,
\end{equation}
to implement $\tilde z$-derivatives via simple matrix multiplication. 
After using a Newton--Kan\-toro\-vich algorithm to linearize the equations of motion for the static and homogeneous background,
we solve them with an LU decomposition. 

The set of equations of motion for the time evolution \eqref{eq:1}-\eqref{eq:4} is solved on a quadratic grid with $512 \times 512$ grid points in the $(x \,\text{-}\,y)$-plane and periodic boundary conditions. 
Derivatives with respect to $x$ and $y$ are implemented via Fourier transforms. 
The equations are then solved as follows.
At every unit of time, we compute the right hand sides of \eqref{eq:1}-\eqref{eq:4}. 
At each $(x, y)$-position we then integrate them with respect to $z$, implementing the boundary conditions \eqref{A0boundary}, \eqref{Axyboundary} and \eqref{Phiboundary} at $z=0$.
This allows us to undo the shifts of the lightcone derivative \eqref{lightconeder} which yields $\partial_t X$, for $X$ again given by $X = A_x, A_y, \Phit$.
These fields can then be propagated in time using a fourth-order/fifth-order Runge–Kutta–Fehlberg algorithm with adaptive timesteps.
The adaptive timesteps vary between $0.001$ and $0.1$ in units defined by $z_\text{h}=1$. 
One unit timestep corresponds to $\Delta t=1$, and is therefore composed of a varying number of numerical timesteps. 
Knowing the fields $ A_x, A_y, \Phit$ at the new time we can plug them into equation \eqref{eq:1} and solve the boundary value problem to also obtain $A_t$ at the new time.
From $\tilde\Phi$ we can then derive $\psi$ according to equation \eqref{bcPhi}.

\section{Vortex Trajectories from the Hall--Vinen--Iordanskii Equations}
\label{VortexTrajCGLE}

The Hall--Vinen--Iordanskii equations \cite{Hall1956a.PRSLA.238.204,Iordansky1964a.AnnPhys.29.335} for the velocities of a system of point vortices 
for vanishing velocity of the normal fluid component are given by
\begin{equation}\label{HVIeq}
	\frac{\text{d}\bm{x}_i}{\text{d}t}=(1-C)\bm{v}^\text{s}_i -w _i C' \bm{e}_\perp\times \bm{v}^\text{s}_i \,,
\end{equation}
where $C, C'$ are phenomenological friction coefficients and $\bm{v}^\text{s}_i$ is the superfluid velocity
at the position of the $i$th vortex of winding number $w_i$ created by all other vortices. We use the unit vector $\bm{e}_\perp$ perpendicular,
in a right-handed coordinate system, to the $(x,y)$-plane. (We point out that ${\bm e}_\perp$ is not the holographic direction
but a third spatial direction only used to write the HVI equations in a concise way.)  

All external forces are set to zero. 
The superfluid velocity at $\bm{x}_i$ is given by 
\begin{equation}
\bm{v}^\text{s}_{i}={2\pi w_{i}}\,{\bm e}_\perp\times \vec{\nabla}_{\bm{x}_{i}} W(\{\bm{x}_k, w_k\})\,,
\end{equation}
where $W(\{\bm{x}_k, w_k\})$ is the Kirchhoff--Onsager functional \cite{Onsager1949a}
\begin{equation}
W(\{\bm{x}_k, w_k|k=1,2\})= 1/(2\pi)\sum_{i\neq j} w_i\,w_j\,\log(|\bm{x}_i - \bm{x}_j|)\,,
\end{equation}
with winding number $w_i$ of the $i$th vortex. 
We further use $\mathbb{I}$ to denote the unit matrix in two-dimensions and introduce $\mathbb{J}= \left(\begin{smallmatrix}
0 & -1 \\
1 & 0 
\end{smallmatrix}\right)$.
Then equation \eqref{HVIeq} can be solved straightforwardly for the case of one vortex $(w_1=1)$ and one anti-vortex $(w_2=-1)$.
Taking the gradient of the Kirchhoff--Onsager functional we obtain 
\begin{equation}
\bm{v}^\text{s}_{i} = 2 \sum_{\{ j|j\neq i\}} w_j \frac{\bm{e}_\perp\times (\bm{x}_i - \bm{x}_j)}{|\bm{x}_i - \bm{x}_j|^2}\,.
\end{equation}

Inserting this into \eqref{HVIeq} leads to 
\begin{equation}
  \label{matrixgl}
\frac{\text{d}\bm{x}_i}{\text{d}t} = 
-2
\begin{pmatrix}
C' & -(1-C)w_i \\
(1-C)w_i &C' \\
\end{pmatrix} 
\frac{\bm{x}_{i} - \bm{x}_{3-i}}{|\bm{x}_1 - \bm{x}_2|^2} 
\end{equation} 
with $i\in\{1,2\}$, and we have used $w_i\,w_{3 -i }=-1$.
We define the vortex--anti-vortex separation 
$\bm{d}=\bm{x}_1 - \bm{x}_2$ with $\bm{d} =d\bm{e}_d$ and the center of mass $\bm{D}=\frac{1}{2}\left(\bm{x}_1 + \bm{x}_2\right)$. 
For the separation we find 
\begin{equation}\label{deom}
\frac{\text{d}\bm{d}}{\text{d}t} = -\frac{4}{d^2} 
\begin{pmatrix}
C' & 0 \\
0 & C' \\
\end{pmatrix}  
\bm{d}=-\frac{4C'}{d^2} \bm{d}\,,
\end{equation}
with $d=|\bm{d}|$, and thus
\begin{equation}
\label{deomshort}
\frac{\text{d}d}{\text{d}t}=\frac{2C'} {d}\,. 
\end{equation}
Similarly, from equation \eqref{matrixgl} we obtain for the center-of-mass motion of the dipole
\begin{equation}\label{Deom}
\frac{\text{d}\bm{D}}{\text{d}t} = \frac{4}{d^2} 
\begin{pmatrix}
0& (1-C) \\
-(1-C)& 0 \\
\end{pmatrix}  
\bm{d}\,,
\end{equation}
which can be rewritten as
\begin{equation}\label{Deomshort}
\frac{\text{d}\bm{D}}{\text{d}t}= \frac{4(1-C)}{d}\,\bm{e}_D\,,
\end{equation}
where we introduced $\bm{e}_D=-\mathbb{J}\bm{e}_d$, the unit vector along the center-of-mass motion. 
Clearly, the center-of-mass motion is orthogonal to the dipole axis, \ie\ $\bm{e}_d \perp \bm{e}_D$. 
Equation \eqref{deomshort} can now be integrated to give 
\begin{equation}\label{dPowerLaw}
d(t) = \sqrt{d(0)^2 - 8 C' t}\,.
\end{equation}
Therefore, the dipole size as function of time follows a square root law. 
Using this, equation \eqref{Deomshort} can be integrated to give
\begin{equation}\label{DofT}
D(t) = D(0) - \frac{1-C}{C'}\left(\sqrt{d(0)^2 - 8 C' t}-d(0)\right)\,.
\end{equation}
From this the solution for $\bm{x}_i$ can be obtained by changing back coordinates. 

Irrespective of the time dependence of the dipole size we can infer directly from the
HVI equation \eqref{HVIeq} that, in the point-vortex limit and for vanishing velocity of the normal
component, the vortices trace out straight lines. In short, one expects straight spatial
trajectories of the vortices when the conditions for the applicability of the HVI equations are met. 

\begin{figure}[t]
	\begin{center}
		\includegraphics{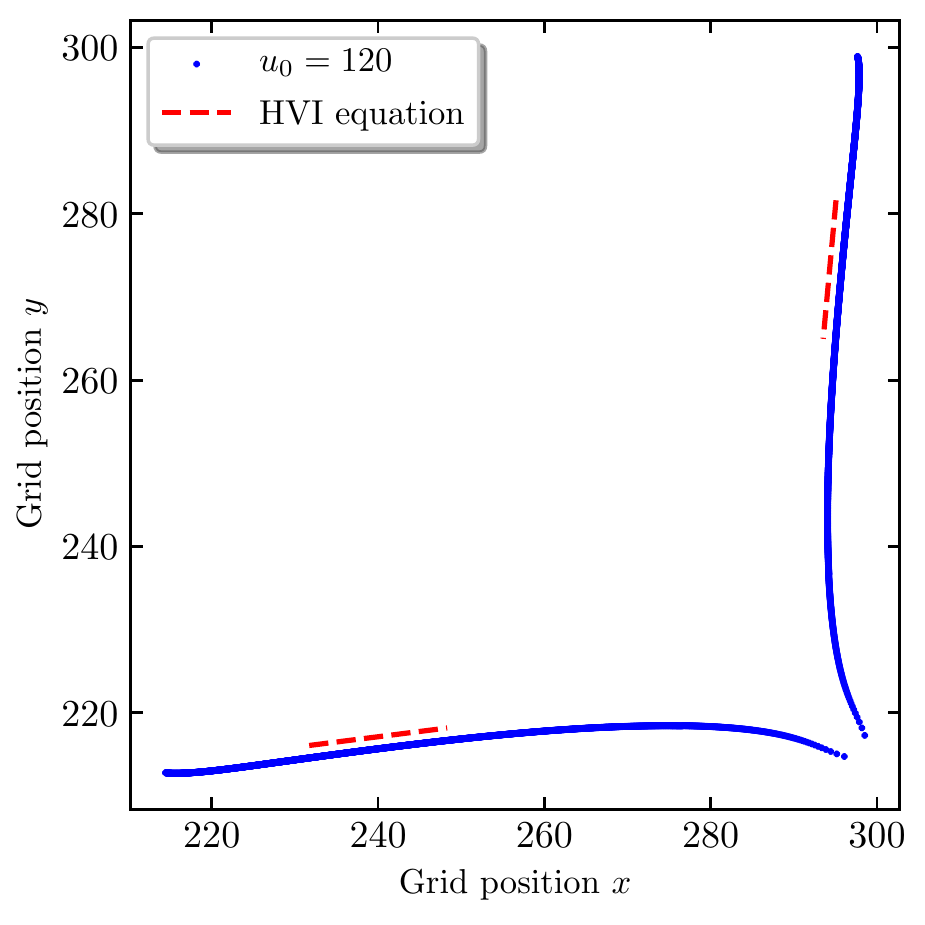}
		\caption{Vortex trajectories for an initial vortex--anti-vortex separation of $u_0=120$ grid points.
                  The dashed red lines indicate the solutions to the HVI equations of motion for the vortex dipole
                  (deliberately shifted by 1.5 grid points). 
                  The time interval for which the trajectories follow straight lines is estimated by eye.
                  At early times, the trajectories are affected by the phase-healing process and consequently do not exactly follow
                  the straight trajectories expected from the HVI equations. 
                  At late times, deviations are expected as the HVI equations only hold in the point-particle approximation
                  which breaks down at smaller separations when the vortices deform and merge. 
		\label{Trajectory120vsHVI}}
	\end{center}
\end{figure}

As we discussed in the main text, a comparison of the holographic vortex trajectories to the HVI equations has two main caveats. 
First, there is the initial phase healing (see section \ref{phasehealing}) which renders the dynamics unphysical in the sense that the system
does not exactly follow the expected behavior of a vortex dipole system for some time after initialization of the two vortices. 
Second, the point-particle approximation of the HVI equation is no longer valid when the vortices come too close to each other. 
Only for an initial separation of $u_0=120$ grid points (or larger) we can clearly identify a segment of the trajectories
for which the vortices follow straight lines, as shown in figure \ref{Trajectory120vsHVI}.
For all other initial separations discussed in the main text the two regimes excluding the application of the HVI equations
seem to overlap and thus prohibit the characteristic behavior to emerge. 
We checked explicitly that for initial separations larger than $u_0=120$ grid points (but small enough for finite-size effects to be negligible) 
the interval for which the trajectories are straight lines increases. This is of course expected due to the universality of the dynamics. 

\section{Dipole Size Power Law}
\label{DipoleSize}

According to the HVI equations discussed in appendix \ref{VortexTrajCGLE},
the vortex--anti-vortex separation for point-like vortices in a superfluid decreases with a power law. 
In the following we check if such a behavior is also observed for the holographic superfluid.
We again exclude from the analysis the early regime in which the phase of the system has not transitioned
(or healed) into that of a proper vortex dipole. 
The phase healing is a dynamical process during the evolution of the system
which is continuous and relatively slow, so that it is difficult to determine its exact endpoint. 
However, using the universality of the dipole dynamics as a criterion, 
it is possible to estimate reasonably well the time after which the system behaves like a proper dipole. 
For that, we can read off from figure \ref{vels} the time when the vortices of a given initial separation follow
the universal (velocity) curve. This time is a good estimate, with an error of about $\Delta t= \pm 30$ timesteps, 
for the endpoint of the phase healing. 

For all times after that endpoint of phase healing we analyze whether the separation $d(t)$ exhibits power-law behavior.
We make the ansatz
\begin{equation}\label{PL}
	d(t)= A(t_0 - t)^b
\end{equation}
for the functional dependence, where $A, t_0$ and $b$ are free parameters, and test how well
it describes $d(t)$ in the considered interval.
In practice we use a Levenberg--Marquardt least-squares fitting algorithm to find the free parameters and an error estimate. 
For the fitting routine we ensure that all points are weighted equally. 

We find that $d(t)$ is indeed well described by the power law \eqref{PL} at intermediate and late times.
For all initial separations we find the same exponent, given by $b=0.53\pm 0.02$, but the parameters $A$ and $t_0$ differ. 
The fitting routine itself yields an error for $b$ of $\Delta b=\pm 10^{-4}$.
However, this is not the true error for the exponent in our analysis as the uncertainty
in the initial time (\ie\ the endpoint of the phase healing process)
used in the fitting routine is much more severe. 
Thus, we vary the initial time by $\Delta t=\pm 30$ to find $\Delta b= \pm 0.02$
which appears to be a more realistic estimate for the error.

\begin{figure}[t]
	\begin{center}
		\includegraphics{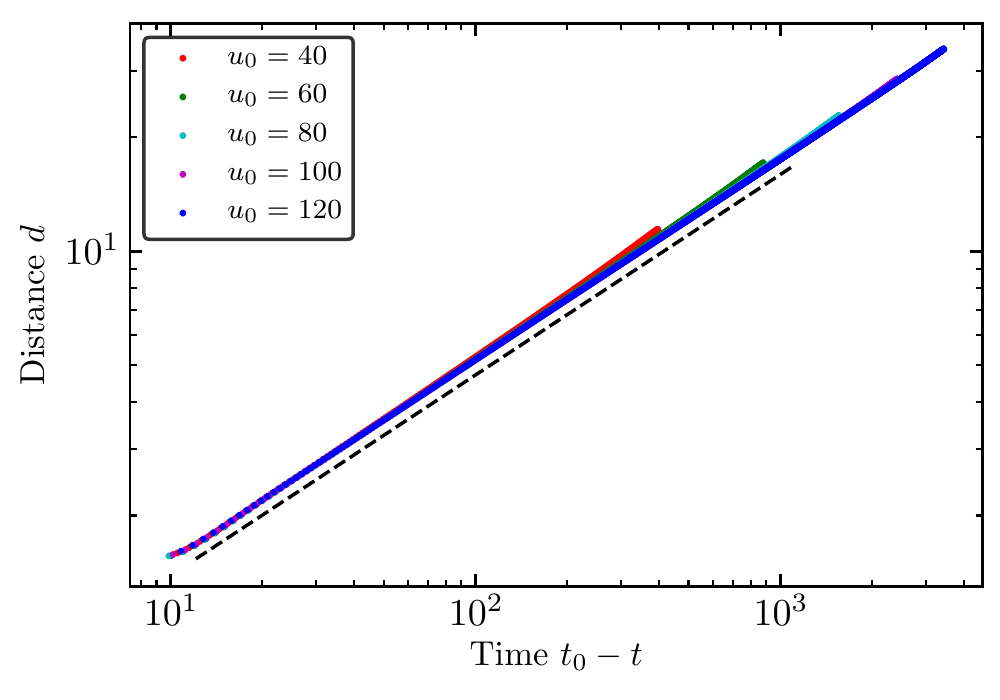}
		\caption{Vortex--anti-vortex separations on a logarithmic scale for all different initial conditions (colors),
                  together with the power law $d(t) \sim (t_0-t)^{0.53}$ (dashed black line). This exponent is obtained, within the error, in 
                  separate fits for all initial distances. For these fits, we exclude the respective parts of the trajectories
                  affected by the phase healing. 
		\label{DistanceLogPowerLaw}}
	\end{center}
\end{figure}

In figure \ref{DistanceLogPowerLaw} we show $d(t)$ on a logarithmic scale for all initial separations. 
Together with the data we also plot as the black line the power law \eqref{PL} with $b=0.53$. 

\section{Finite Difference Methods}
\label{fdm}

To compute the velocities and accelerations from the trajectories of the vortex--anti-vortex system we use finite difference methods. 
Finite differences allow for a good approximation of the derivatives of our quasi-continuous vortex trajectories.

As we use an adaptive time-stepping scheme, our timesteps are unevenly spaced. 
This has to be taken into account when computing finite differences. 
Nevertheless the calculation is straightforward. For simplicity, we restrict ourselves here to one spatial dimension in
the following, the generalization to two dimensions being obvious. 
We start out by Taylor expanding the position of one of the vortices at times $t_{i+1}$ and $t_{i-1}$, 
\begin{align}
x(t_{i+1})&=x(t_i+\Delta t)=x(t_i)+  \dot{x}(t_i)\,\Delta t + \ddot{x}(t_i)\,\frac{(\Delta t)^2}{2} + \ldots\label{tp}\,,\\
x(t_{i-1})&=x(t_i-\Delta \tilde{t})=x(t_i)-  \dot{x}(t_i)\,\Delta \tilde{t} + \ddot{x}(t_i)\,\frac{(\Delta \tilde{t})^2}{2} + \ldots\label{tm}\,,
\end{align}
where a dot denotes the derivative with respect to the time $t$ and we introduced both $\Delta t=t_{i+1} - t_i$
and $\Delta \tilde{t}=t_i - t_{i-1}$ to account for the in general not evenly spaced timesteps.
Computing the difference $\eqref{tp}-\eqref{tm}$ and neglecting all terms of second or higher order in the derivative gives
the local velocity of the vortex at time $t_i$, 
\begin{equation}
v(t_i)=\dot{x}(t_i)=\frac{1}{2}\left(\frac{x(t_{i+1})}{t_{i+1}-t_{i}} - \frac{x(t_{i-1})}{t_{i}-t_{i-1}}\right). 
\end{equation}

We are also interested in the vortex acceleration.
For that, we compute the sum $\eqref{tp} + \eqref{tm}$  and neglect all terms of third or higher order in the derivative. 
We obtain 
\begin{align}
\frac{x(t_{i+1})}{\Delta t}+\frac{x(t_{i-1})}{\Delta \tilde{t}}=\frac{x(t_i)}{\Delta t}+ \frac{x(t_i)}{\Delta \tilde{t}}+\frac{1}{2}\left(\ddot{x}(t_i)\Delta t +\ddot{x}(t_i)\Delta\tilde{t}\right)\,,
\end{align}
which, after some algebra, gives us the central second finite differences derivative used in this work,
\begin{align}
a(t_i) = \ddot{x}(t_i)=\frac{2}{t_{i+1}-t_{i-1}}\left[\frac{x(t_{i+1})}{t_{i+1}-t_i}+\frac{x(t_{i-1})}{{t_{i}-t_{i-1}}}-\frac{x(t_{i})}{t_{i+1}-t_i}-\frac{x(t_{i})}{{t_{i}-t_{i-1}}}\right]\,.
\end{align}

\providecommand{\href}[2]{#2}\begingroup\raggedright

\end{document}